\documentclass[aps,prc,twocolumn,superscriptaddress]{revtex4-1}
\usepackage{graphicx}
\usepackage{amsmath}

\usepackage[colorlinks,
linkcolor=blue,
anchorcolor=blue,
urlcolor=red,
citecolor=blue]{hyperref}

\begin{document}

%\preprint{}

%\title{Importance of continuum coupling and three-body forces on unbound neutron-rich oxygen isotopes in the Gamow shell model}

\title{Unbound spectra of neutron-rich oxygen isotopes predicted by the Gamow shell model}


\author{J. G. Li}
\affiliation{School of Physics,  and   State Key  Laboratory  of  Nuclear  Physics   and  Technology, Peking University, Beijing  100871, China}
\author{N. Michel}
\affiliation{Institute of Modern Physics, Chinese Academy of Sciences, Lanzhou 730000, China}
\affiliation{School of Nuclear Science and Technology, University of Chinese Academy of Sciences, Beijing 100049, China}
\author{W. Zuo}
\affiliation{Institute of Modern Physics, Chinese Academy of Sciences, Lanzhou 730000, China}
\affiliation{School of Nuclear Science and Technology, University of Chinese Academy of Sciences, Beijing 100049, China}
\author{F. R. Xu}\email[]{frxu@pku.edu.cn}
\affiliation{School of Physics,  and   State Key  Laboratory  of  Nuclear  Physics   and  Technology, Peking University, Beijing  100871, China}


\date{\today}

\begin{abstract}

The Gamow shell model has shown to efficiently describe weakly bound and unbound nuclear systems, as internucleon correlations and continuum coupling are both taken into account in this model.
In the present work, we study neutron-dripline oxygen isotopes. 
It is hereby demonstrated that the presence of continuum coupling  is important for the description of oxygen isotopes at dripline, and especially to assess the eventual bound or unbound character of $^{28}$O.
Our results suggest that the ground state of $^{28}$O is weakly unbound and is similar to the narrow resonant $^{26}$O ground state.
Predictions of weakly bound and resonance excited states in $^{24\text-26}$O are also provided.
The asymptotes of the studied many-body states are analyzed via one-body densities, whereby the different radial properties of well bound, loosely bound, resonance states are clearly depicted.

\end{abstract}

%\keywords{}

\maketitle

\section{Introduction}

The study of nuclei near the dripline is one of the most important topics of interest in current nuclear physics.
The comprehension of the properties of exotic nuclei near driplines and of the stability of nuclei in general is a challenge for nuclear theory.
Neutron-rich oxygen isotopes are particularly interesting for that matter.
Due to the presence of atypical features, such as halos, resonance ground states, and fairly broad low-lying unbound states,
neutron-rich oxygen isotopes offer an important testing ground to understand the structure of neutron-rich nuclei in extreme conditions,
insofar as both their experimental and theoretical studies can be thoroughly performed.

$^{22,24}$O have doubly magic nature for the neutron number \textit{N} = 14, 16 \cite{PhysRevLett.96.012501, PhysRevC.69.034312, PhysRevLett.100.152502, PhysRevLett.109.022501}.
Experiments have shown that $^{25,26}$O are unbound \cite{PhysRevLett.116.102503}, which suggests that $^{24}$O is the heaviest bound isotope of the oxygen chain \cite{PhysRevLett.110.152501,PhysRevLett.116.102503}.
The ground states of $^{25,26}$O are unbound by about 835 and 18 keV with respect to the ground state of $^{24}$O, respectively \cite{PhysRevLett.116.102503}.
The barely unbound property of the ground state of $^{26}$O is a strong incentive to ponder about the bound or unbound character of $^{28}$O.
Several low-lying states in $^{25\text-26}$O have been reported in experiments.
In $^{25}$O, $1/2^+$ excited state, located at 3.3 MeV above the ground state, is suggested in Ref. \cite{PhysRevC.96.054322}.
Experiments have shown that the ground state of $^{26}$O exhibits two-neutron decay \cite{PhysRevLett.110.152501,PhysRevLett.116.102503}.
An unbound excited $2^+$ state in $^{26}$O has also been detected experimentally \cite{PhysRevLett.116.102503}.
Clearly, a proper theoretical description of this $2^+$ state in $^{26}$O is a challenge for theoretical models \cite{PhysRevLett.116.102503}.

Contrary to well-bound nuclei which can be treated as closed quantum systems, dripline nuclei are open quantum systems, as they can be weakly bound or unbound with respect to nucleon emission.
Continuum coupling has been shown to play an important role in loosely bound and unbound nucleonic systems \cite{PhysRevLett.89.042501,PhysRevLett.89.042502}.
The proper treatment of continuum coupling has always been a challenging problem for theoretical methods \cite{1402-4896-2013-T152-014022}.
The Gamow shell model (GSM) \cite{PhysRevLett.89.042501,PhysRevLett.89.042502,0954-3899-36-1-013101} is an appropriate tool for that matter.
GSM allows to describe many-body bound states and resonances within a unique framework, where both internucleon correlations and continuum coupling are included.
Continuum coupling arises from the use of the Berggren ensemble containing bound, resonance, and continuum one-body states, so that it is present at the basis level \cite{BERGGREN1968265}.
GSM has been successfully applied to many situations of physics interests. For example, one can mention the halo structures of $^{6,8}$He \cite{PhysRevC.84.051304} and $^{11}$Li \cite{PhysRevC.74.054305}, the spectra and correlation densities of $psd$-shell neutron-rich nuclei \cite{PhysRevC.96.054316}, neutron-rich oxygen isotopes \cite{SUN2017227,PhysRevC.96.054322,PhysRevC.96.024308}, and proton-rich isotonic systems of $^{16}$O \cite{PhysRevC.100.064303}, where the use of  effective field theory (EFT) \cite{MACHLEIDT20111,PhysRevC.98.044301} to generate the residual nucleon-nucleon interaction in GSM has been introduced.

Several models have been applied to describe neutron-rich oxygen isotopes \cite{PhysRevLett.108.242501,PhysRevC.80.051301,PhysRevLett.105.032501,PhysRevLett.113.142501,PhysRevLett.113.142502,Holt2013,PhysRevC.96.024308,SUN2017227,PhysRevLett.118.032502,Holt:2019,PhysRevC.99.061302,HU2020135206,MA2020135257}. \textit{Ab initio} calculations \cite{PhysRevLett.108.242501,PhysRevLett.113.142501,PhysRevLett.113.142502,PhysRevLett.118.032502,SUN2017227,Holt:2019,PhysRevLett.105.032501,Holt2013,HU2020135206,MA2020135257} with  the in-medium similarity renormalization group (IM-SRG) \cite{PhysRevLett.118.032502,Holt:2019,PhysRevC.99.061302}, coupled-cluster (CC) \cite{PhysRevLett.108.242501,PhysRevLett.113.142502} and many-body perturbation theory (MBPT) \cite{Holt2013,SUN2017227,MA2020135257} have shown that three-body forces should be included to describe the binding energy saturation of the heaviest oxygen isotopes.
However, these calculations bear a large theoretical uncertainty in $^{18\text -26}$O. This renders comparison with experimental data difficult,
and accurate predictions can be hardly made in neutron-rich oxygen isotopes.
In addition, results arising from \textit{ab initio} calculations depend on realistic nuclear forces used (a short summary of the IM-SRG calculations based on different chiral nuclear forces can be found in Ref. \cite{PhysRevC.96.024308}).
Continuum coupling effect is absent in the IM-SRG  calculations \cite{PhysRevLett.113.142501,PhysRevLett.118.032502,Holt:2019} and shell model calculations \cite{PhysRevC.74.034315,PhysRevC.85.064324,PhysRevC.98.044301}  based on phenomenological effective interactions.
The inclusion of continuum effect within the continuum shell model (CSM) \cite{PhysRevLett.94.052501} allows to predict that  $^{28}$O is unbound.
The GSM calculations using the Furutani-Horiuchi-Tamagaki (FHT) effective nuclear force \cite{FHT1,FHT2} give loosely bound ground states for $^{26}$O and $^{28}$O \cite{PhysRevC.96.024308}.
However, the theoretical uncertainties are of the order of 500 keV in the calculations of Ref. \cite{PhysRevC.96.024308}, which is comparable to neutron separation energy in $^{25\text-28}$O.
It is then difficult to make predictions for oxygen isotopes at neutron dripline using the FHT Hamiltonian, which is fitted with experimental data \cite{PhysRevC.96.024308}.
The same phenomenon is found in the proton dripline isotonic systems of $^{16}$O \cite{PhysRevC.100.064303},
where the use of EFT to generate the nucleon-nucleon interaction could provide with a better reproduction of experimental data compared to the FHT framework.
Consequently, we will use EFT to devise the Hamiltonian to generate the nuclear states of neutron-dripline oxygen isotopes in this paper.

This paper is organized as follows.
The GSM basis features, the used model space and Hamiltonian, and the optimization procedure to find the parameters of the Hamiltonian will be described in Sec. \ref{Method}.
In particular, we will emphasize the use of EFT and $A$-dependence in the nucleon-nucleon interaction.
The effect of three-body forces will also be discussed in neutron-rich isotopes.
We will then report and discuss our results in Sec. \ref{Results}. They consist of the spectra of oxygen isotopes and of one-body densities of the associated nuclear states.
In particular, the four-neutron separation energy of $^{28}$O will be assessed in our calculation.
Low-lying resonances are predicted in $^{23\text-28}$O and the properties of those states will be analyzed from the one-body densities of their valence neutrons.

\section{Method} \label{Method}

GSM is a multiconfiguration shell model framework \cite{PhysRevLett.89.042501,PhysRevLett.89.042502,0954-3899-36-1-013101} based on the use of the one-body Berggren basis \cite{BERGGREN1968265}.
The Berggren basis consists of bound, resonance, and scattering one-body states generated by a finite-depth potential \cite{BERGGREN1968265}, with a completeness relation for each partial wave,
\begin{eqnarray}\label{equation1} 
   \sum_{n} |\phi_{n}\rangle\langle \widetilde {\phi}_{n}| + \int_{L_{+}}|\phi(k)\rangle\langle \widetilde {\phi}(k)| dk = 1,
\end{eqnarray}
where $|\phi_{n}\rangle$ is a bound or resonance state, $|\phi(k)\rangle$ are continuum states considered along a contour in the complex plane, denoted as $L^+$. The tilde sign in Eq. (\ref{equation1}) indicates that we use the Berggren metric, allowing for complex-energy eigenvalues \cite{BERGGREN1968265}.
The $L^+$ contour starts from $k = 0$, peaks at a given $k_p$ value in the lower complex plane, goes back to the real $k$ axis in $k_m$ and extends to infinity afterwards, as illustrated in Fig. 1 of Ref. \cite{SUN2017227}.
The $L^+$ contour has to encompass the resonance states of the finite sum of Eq. (\ref{equation1}) \cite{BERGGREN1968265}.
Both bound and resonance states can be expanded using Eq. (\ref{equation1}) \cite{BERGGREN1968265}.
The $L^+$ contour is truncated at $k = k_{\text{max}}$ on the real $k$ axis.
In the present paper, we take a large enough $L^+$ momentum truncation of $k_{\text{max}}=5$ fm$^{-1}$ for the calculations of neutron-rich oxygen isotopes.
The $L^+$ contour of Eq. (\ref{equation1}) must be discretized in practical applications.
For this, the Gauss-Legendre quadrature has proved to be the most efficient discretization scheme,
as numerical precision is achieved with 30$-$50 discretized states \cite{PhysRevLett.89.042502,PhysRevLett.89.042501,SUN2017227}.
Therefore, GSM provides with continuum coupling at basis level, and inter-nucleon correlations are present via configuration mixing.
GSM is then the tool of choice to describe multinucleon systems of complex structure at driplines, where continuum degrees of freedom are prominent.

We will work in the  picture of a core plus valence neutron.
For this, the closed-shell nucleus $^{22}$O is selected to be the inner core.
The one-body potential is mimicked by a Woods-Saxon (WS) potential, whose parameters are adjusted to reproduce the single-particle spectrum of $^{23}$O \cite{ensdf}.
The two-body interaction is of the pionless EFT character \cite{MACHLEIDT20111}, to which an $A$-dependence is added in order to account for the effect of three-body force.
Similar methods have been used for the description of $sd$-shell nuclei with real-energy shell model  \cite{PhysRevC.74.034315} 
and proton-rich nuclei in the $A \approx$ 20 region with GSM \cite{PhysRevC.100.064303}.
Owing to the few data that can enter the optimization procedure, only the leading-order (LO) nucleon-nucleon interaction of the EFT interaction \cite{CONTESSI2017839,RevModPhys.85.197,RevModPhys.92.025004}  are used in the present work.  In then pionless EFT framework \cite{CONTESSI2017839,RevModPhys.85.197,RevModPhys.92.025004},
the LO nucleon-nucleon interaction in momentum space reads,
\begin{eqnarray}\label{equation2}
  \langle \boldsymbol{p^{\prime}}  |V_{NN}^{\rm LO} | \boldsymbol{p} \rangle = C_S + C_T \left( \boldsymbol{\sigma}_1 \cdot \boldsymbol{\sigma}_2 \right),
\end{eqnarray}
where the  $\boldsymbol{p^{\prime}}$ and $\boldsymbol{p}$  are the outgoing and incoming relative momenta, respectively, and 
the interaction only contains spin-independent and spin-dependent terms. 
We can separate the interaction into a spin singlet channel $^1S_0$ and a spin triplet $^3S_1$, as follows
\begin{eqnarray}\label{equation2.2}
  \langle \boldsymbol{p^{\prime}}  |V_{NN}^{\rm LO} | \boldsymbol{p} \rangle  = C_{S0} \frac{1-P_\sigma}{2} + C_{S1} \frac{1+P_\sigma}{2},
\end{eqnarray}
where $P_\sigma = \frac{1}{2}(1+\boldsymbol{\sigma}_1 \cdot \boldsymbol{\sigma}_2)$ is the spin exchange operator. $C_{S0} = (C_S-3C_T)$ and $C_{S1}= (C_S+C_T)$ are interaction strengths for the $^1S_0$ and $^3S_1$ channels, respectively.
%In terms of partial waves of nucleon-nucleon scattering phase shifts, the LO term on the $S$ partial waves produces two interaction strengths: $V_{\text{cont}}^{\text{LO}} (^1S_0) = 4 \pi (C_S-3C_T)$ for the spin singlet channel $^1S_0$, and $V_{\text{cont}}^{\text{LO}} (^3S_1) = 4 \pi (C_S+C_T)$ for the spin triplet channel $^3S_1$ \cite{MACHLEIDT20111}.
In the present GSM calculations of neutron-rich oxygen isotopes with the $^{22}$O core, the active valence nucleons are only neutrons. Due to the Pauli exclusion principle, the $^3S_1$ channel should not be present. Therefore, only the $^1S_0$ channel contributes to the interaction, and only the strength $C_{S0}$ needs to be determined (by fitting available experimental data of the studied oxygen isotopes). 

The EFT interaction is often renormalized by way of a momentum-dependent regulator function, see Ref. \cite{MACHLEIDT20111} for details. Recently, in Refs. \cite{PhysRevC.93.044332,PhysRevC.98.044301,PhysRevC.98.054301}, it was shown that the EFT interactions can be regularized in the HO basis within a limited model space, and the approach gave converged calculations for heavier nuclei. Moreover, the approach automatically cutoff the high-momentum parts, and therefore no additional momentum-dependent regulator function is necessary. The similar treatment has been employed in the recent works to renormalize the EFT interaction \cite{PhysRevC.93.044332,PhysRevC.86.031301,PhysRevC.90.064007,PhysRevC.98.054301,PhysRevC.98.044301}. In practice, one has to choose the oscillator frequency $\hbar w$ and the considered maximum shell number $N_{\text{max}}$ of the momentum-space HO basis. Then, the cutoff is given by \cite{PhysRevC.90.064007,PhysRevC.86.031301,PhysRevC.93.044332,PhysRevC.98.054301}
\begin{eqnarray}\label{equation3}
  \Lambda = \sqrt{2(N_{\text{max}}+2+3/2)}\hbar/b,
\end{eqnarray}
where $b$ is oscillator length, $b\equiv \sqrt{\hbar/(\mu w)}$, and the $\mu$ is the reduced mass of the two-nucleon system.
To include the effects of three-body force, an $A$-dependent two-body factor is multiplied to the two-body matrix elements of the EFT interaction.
The $A$-dependent two-body factor is standard and reads as \cite{Bohr_Mottelson}
\begin{eqnarray}\label{equation5}
   F_{2b}= \left(\frac{A_{\rm core}+2}{A}\right)^e,
\end{eqnarray}
where $A$ is the number of nucleons of the nucleus, $A_{\rm core}$ is the number of nucleons of the inner core, and \textit{e} is the exponent parameter.

In the GSM calculations, one can choose the \textit{sd} partial waves as the active model space for valence neutrons outside the $^{22}$O core. The parametrized interaction strength is then dependent on the model space chosen. Though effects from higher partial waves (e.g., the \textit{pf} partial waves) may be included through the parametrization of the interaction, we will see that the inclusion of the \textit{pf} partial waves in the active space makes sense for some cases of unbound isotopes beyond the dripline. For example, in $^{28}$O, the $N=20$ Fermi surface is over the  $0d_{3/2}$ orbital, and hence configurations with neutrons occupying the  \textit{sd} and \textit{pf} continua should be considered as active components in  shell model calculations. In the present work, we will mainly choose the \textit{sdpf} space for the GSM calculations, but the calculations using the \textit{sd} space will be also displayed for comparison. 
As we will see, the energy differences between calculations using the \textit{sd} and \textit{sdpf} spaces are indeed small.
Due to the high centrifugal barrier of the partial waves bearing $\ell > 3$, their influence on wave function asymptotes is negligible.

In the Berggren basis with the $^{22}$O core, the $1s_{1/2}$ orbital is bound, and $0d_{3/2}$ is a resonance, while the $p$ and $f$ partial waves are nonresonant scattering continua. 
However, due to the high centrifugal barrier of $f$ partial waves, their components in the many-body wave functions should be localized inside the nucleus. In addition, contrary to the \textit{s} and \textit{d} channels, the \textit{f}  partial waves have no resonant poles.
Therefore, it should be reasonable that the $f$ partial wave is represented within the harmonic oscillator (HO) basis, to reduce the model space dimension and computational task. The \textit{sdp} partial waves are expanded in the Berggren basis.
The Berggren basis contours for the $sdp$ partial waves are defined with $k_p = (0.15,-0.10)$, $k_m = (0.30,0)$, and $k_{\text{max}} = (5.0,0)$ fm$^{-1}$ in the complex-momentum space (see above for notations).
Each segment is discretized by ten points along with the $L^+$ contour.
The $f$ partial waves are represented by six HO shells.
It has been checked that the used representation for \textit{sdpf} partial waves is sufficient to give converged results for the considered nuclei.

Due to the explosive growth of the model space dimension in GSM, it is necessary to truncate the GSM model space by limiting the number of occupied scattering states.
In fact, we can only have two neutrons at most in the nonresonant continuum (denoted as 2p-2h truncations) to obtain tractable dimensions when using  discretized Berggren basis contours.
To reduce the computational cost and obtain  better convergences, we use natural orbitals \cite{NaturalOrbitals} as one-body states instead of Berggren basis states.
Natural orbitals are the eigenstates of the scalar density matrix of the considered many-body state \cite{NaturalOrbitals}.
Therefore, they recapture a large part of the strength of many-body systems,
so that the occupation of natural orbitals by valence nucleons decreases very quickly with the radial quantum number of natural orbitals.
Recent applications of natural orbitals in GSM showed that one typically needs three to five natural orbitals per partial wave,
comparing with the 30 discretized states needed at least when using a Berggren basis contour \cite{PhysRevC.96.054316,PhysRevC.100.054313}.
Hence, the GSM model space dimension is much reduced, so that it is possible to perform calculations with 3p-3h truncations, whereby convergence is well obtained.
The scalar density matrix of the many-body state must clearly be calculated in a GSM truncated space.
The GSM model space defined with 2p-2h truncations has shown to provide with almost converged scalar density matrices.
As a consequence, by using the described two-step calculation, one could achieve convergence in the GSM many-body space while diagonalizing a GSM Hamiltonian matrix of tractable dimension.

\begin{table}[!htb]
\centering
\caption{\label{Table.InterParamEFT} The optimized strength $C_{S0}$ (in 10$^{-2}$ MeV$^{-2}$ \cite{MACHLEIDT20111}) of the EFT LO interaction using different momentum cutoffs $\Lambda$ (in MeV) in GSM calculations within the \textit{sdpf} and \textit{sd} model spaces. The different momentum cutoffs are obtained  in Eq. (\ref{equation3}) from varying the oscillator frequency $\hbar w$ (in MeV) in a model space with $N_{\text{max}} = 10$.}
\setlength{\tabcolsep}{2.96mm}{
\begin{tabular}{lccccc} \hline \hline
$\hbar w$ & 8    & 10   &  12 & 15 \\  \hline
$\Lambda$ &  318   & 356  & 390 & 436 \\  \hline
 \textit{sdpf}  & $-0.0572$ &   $-0.0492$   & $-0.0466$ & $-0.0449$ \\  \hline
  \textit{sd}   & $-0.0709$ &    $-0.0645$  &  $-0.0621$  &  $-0.0620$\\  \hline \hline
\end{tabular}}
\end{table}

The strength of the EFT interaction is optimized to reproduce the binding energies of $^{24\text-26}$O \cite{ensdf}.
For the WS potential of the $^{22}$O core, we fix the diffuseness $a$ = 0.65 fm, the radius $R_0$ = 3.15 fm, and the spin-orbit strength $V_{ls}$ = 7.5 MeV, only the $V_0$ is adjusted by fitting the experimental single-particle spectrum of $^{23}$O \cite{ensdf}.
The $\ell$ dependence of the  fitted central strength $V_0$ reads $V_0$ = 51.33 MeV if $\ell = 0$ and $V_0$ = 52.60 MeV if $\ell \geq 1$.

The EFT interaction often depends on a cutoff in momentum space.
To assess this dependence, we will consider different momentum cutoffs $\Lambda$ obtained from varying the oscillator frequency $\hbar w = 8, 10, 12$, and $15$ MeV  in a model space with $N_{\text{max}}= 10$, corresponding to the $\Lambda$ = 318, 356, 390, and 436 MeV, respectively, whose interactions are labeled by EFT(318), EFT(356), EFT(390), and EFT(436), respectively.
In order to proceed with the optimization procedure of binding energies in $^{24\text-26}$O, we firstly devise $A$-independent EFT(318), EFT(356), EFT(390), and EFT(436) interactions. The optimized strength of $C_{S0}$ of the EFT interaction is shown in Table. \ref{Table.InterParamEFT}. Binding energies are slightly overbound using EFT(318),
whereas the calculations using EFT(356), EFT(390), and EFT(436) well reproduce the binding energies of $^{24\text-26}$O (see Fig. \ref{Binding_O24}). 
Thus, $A$-dependent two-body factors are needed for EFT(318).
We will then multiply the initial strength of $C_{S0}$ in Table. \ref{Table.InterParamEFT} obtained with EFT(318) by an $A$-dependent two-body factor.
We adjusted the $A$-dependent two-body factor in order to reproduce the two-neutron separation energy of the $^{26}$O.
We obtained that the exponent parameter appearing in Eq. (\ref{equation5}) is $e = 0.4$, which is consistent with that in Refs. \cite{PhysRevC.100.064303,PhysRevC.98.044301,PhysRevC.74.034315}. It may be understood that a soft interaction usually needs a modification from the effect of three-body interaction.

\section{Results} \label{Results}

Continuum coupling starts to be an important part of neutron-rich oxygen isotopes heavier than $^{24}$O.
While $^{24}$O is doubly magic, being mostly built from the well-bound $(1s_{1/2})^2$ configuration above the $^{22}$O core \cite{PhysRevLett.109.022501},
oxygen isotopes beyond $^{24}$O mainly occupy configurations involving the unbound $0d_{3/2}$ orbital \cite{PhysRevLett.108.142503,PhysRevLett.116.102503}.
The ground state of $^{26}$O has been shown to decay via dineutron emission, with an energy just about 18 keV above the ground state of $^{24}$O \cite{PhysRevLett.116.102503}.
$^{26}$O is thus slightly unbound, then directly pointing out the question whether $^{28}$O is bound or not, which is, in fact, still open \cite{PhysRevLett.110.152501,PhysRevC.96.024308}.

\begin{figure}[!htb]
\includegraphics[width=1.00\columnwidth]{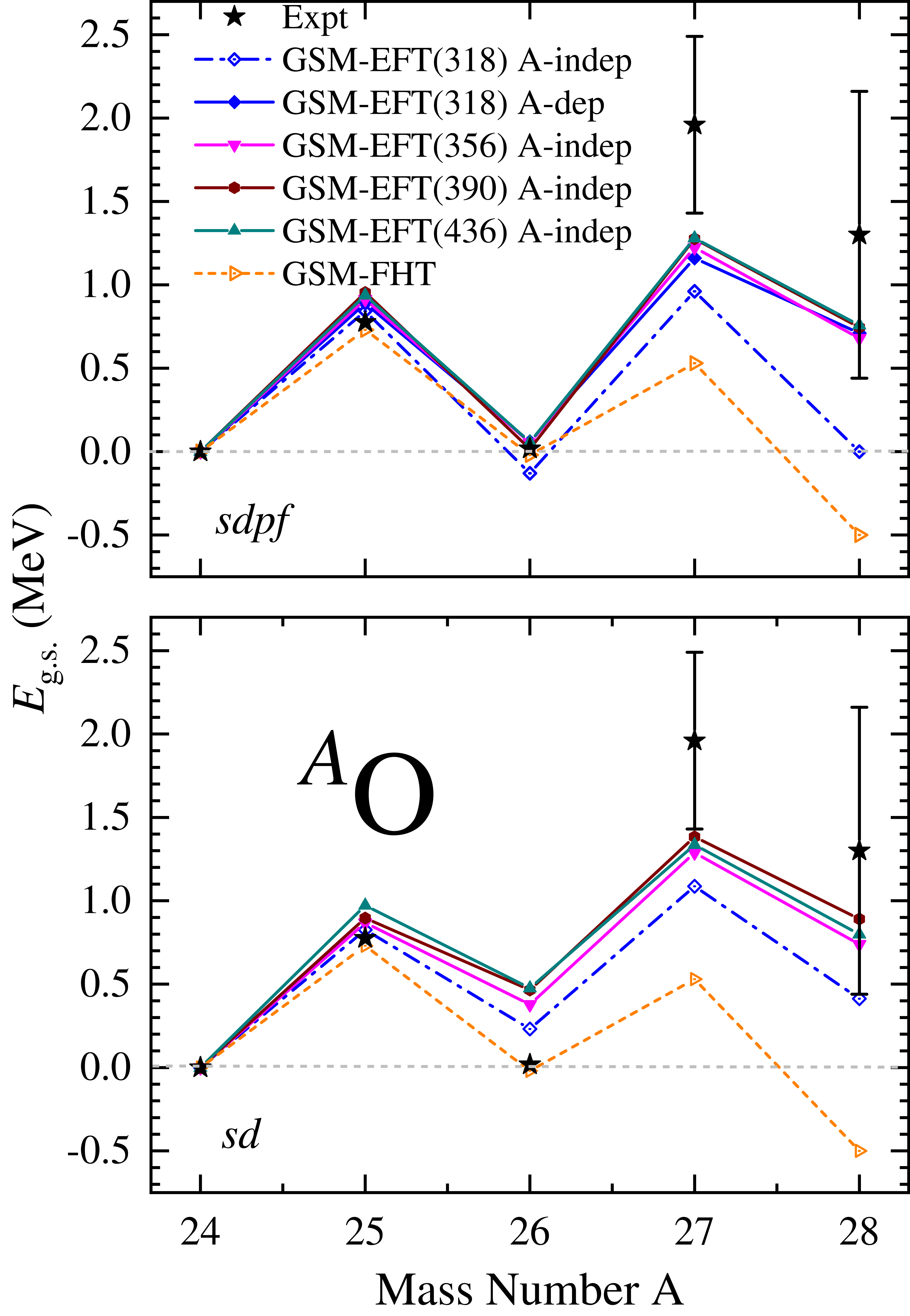}% Here is how to import EPS art
\caption{Energies of ground states in $^{24\text-28}$O, calculated by GSM within the \textit{sdpf} (upper) and \textit{sd} (lower) model spaces, using the EFT(318), EFT(356), EFT(390), EFT(436),  and FHT interactions, with $A$-independence ($A$-indep) or $A$-dependence ($A$-dep) (see text for definitions). Results are compared with data available \cite{PhysRevLett.116.102503,wang2017ame2016},  represented by a star. The data for $^{25,26}$O and  $^{27,28}$O are taken from the experiment \cite{PhysRevLett.116.102503} and evaluations given in AME2016 \cite{wang2017ame2016}, respectively. }{\label{Binding_O24}}
\end{figure}

We calculated the energies of the ground states of $^{24\text{-}28}$O with GSM within \textit{sdpf} active space, using the fitted EFT(318), EFT(356), EFT(390), and EFT(436) interactions (see Sec. \ref{Method}).
The obtained $\chi^2$ deviation in our GSM calculations is about 100 keV for all considered interactions, so that the fitted Hamiltonian can be deemed to be optimal. GSM calculations using the EFT(318), EFT(356), EFT(390), and EFT(436) interactions are depicted in Fig. \ref{Binding_O24} along with  available data \cite{PhysRevLett.116.102503,wang2017ame2016} and the GSM calculations based on a fitted FHT interaction using  $sdp$ partial waves in Ref. \cite{PhysRevC.96.024308}.

GSM calculations with $A$-independent EFT(318) and FHT interactions both provide a slightly overbound ground state for $^{26}$O.
The energy of the ground state of $^{26}$O is too bound by 200 keV using the EFT(318) interactions and by 30 keV using the FHT interaction, compared with experimental data \cite{PhysRevLett.116.102503}.
However, the energy of the ground state of $^{26}$O becomes more bound with the FHT interaction if one includes the $f$ partial waves in the GSM model space \cite{PhysRevC.96.024308}.
Moreover, when using the FHT interaction, the obtained ground-state energy of $^{24}$O is less bound by 600 keV, compared with experimental data.
Consequently, if one makes the FHT interaction more binding in order to reproduce the experimental energy of the ground state of $^{24}$O,
the $^{26}$O ground state will depart even more from experimental data.
The tendency of the FHT interaction to overbind many-body nuclear systems had also been noticed when considering the proton-rich isotones of $^{16}$O \cite{PhysRevC.100.064303}.
In the absence of genuine three-body force in the Hamiltonian, overbinding can only be counterbalanced by using $A$-dependent Hamiltonians, see Eq. (\ref{equation5}).
Indeed, $A$-dependent EFT(318) interactions reproduce experimental data well using this approach (see Fig. \ref{Binding_O24}).
Note that an $A$-dependent Hamiltonian using an EFT interaction could also successfully describe the proton-rich isotones of $^{16}$O \cite{PhysRevC.100.064303}.
The GSM calculations using EFT(356), EFT(390), and EFT(436) without $A$-dependence provide a good description of the experimental data associated to neutron-rich oxygen nuclei.
Therefore, the $A$-dependent EFT(318) and the $A$-independent EFT(356), EFT(390), and EFT(436) interactions provide good agreements of the $^{23\text -26}$O ground states with experimental data,
where, in particular, the two-neutron separation energy of $^{26}$O is about 20 keV.
The calculated ground state of $^{28}$O is unbound in all three cases and located about 700 keV above the ground state of $^{24}$O.
The unbound character of $^{28}$O obtained in our calculations suggests that the neutron dripline  of oxygen  is located at $^{24}$O.

In Fig. \ref{Binding_O24}, we also show the calculations using only the \textit{sd} partial waves as the active  shell model space. The interaction strength of $C_{S0}$ is then refitted, as shown in Table \ref{Table.InterParamEFT}. 
%\textbf{ The obtained $\chi^2$ deviation is about 150 keV, which is a little large than the above results obtained within \textit{sdpf} active space.}
It is seen that the energy differences between calculations using the \textit{sdpf} and \textit{sd} spaces are indeed small. All the calculations within $sd$ spaces provide that the $^{25\text{-}28}$O are unbound, which are close to experimental data and calculations within $sdpf$ space. However, the calculations using $sd$ space provide that the ground state of $^{26}$O is unbound about 300 keV, a little higher than experimental value which is about 20 keV unbound \cite{PhysRevLett.116.102503}. Though the energy difference obtained using the two different spaces is small, the calculation within $spdf$ space seems to be more reasonable.

After pondering about the theoretical aspects of our Hamiltonians, we now compare our results with other nuclear models.
Many theoretical works \cite{PhysRevLett.105.032501,PhysRevLett.113.142502,PhysRevLett.94.052501,Holt:2019,Holt2013} have been done to explain the properties of the oxygen isotopes at neutron dripline.
\textit{Ab initio} IM-SRG calculations \cite{Holt:2019,PhysRevLett.118.032502} lead to the conclusion that the $^{25\text -28}$O are unbound, whereas the theory-experiment difference  on the two-neutron separation energy of $^{26}$O is about 2 MeV.
$^{26}$O is slightly bound by about 500 keV with respect to $^{24}$O, and $^{28}$O is unbound in realistic shell model calculations \cite{Holt2013}.
In shell model calculations using the effective USDB \cite{PhysRevC.74.034315} and YSOX \cite{PhysRevC.85.064324} interactions,
both $^{26,28}$O isotopes are unbound and the theory-experiment differences  on the binding energy of $^{26}$O are about 0.3 and 1 MeV in USDB and YSOX calculations, respectively, when compared to experimental data.
However, the above calculations did not take into account continuum coupling.
CC calculations \cite{PhysRevLett.108.242501} including continuum coupling predicted that $^{25,26}$O are unbound about 400 keV and 100 keV with respect to $^{24}$O, respectively.
CC results are then in good agreement with experimental data.
$^{28}$O is found to be unbound with respect to the $^{24}$O about 4 MeV in CC calculations.
However, the experimental energy differences between light and heavy oxygen isotopes are not correctly reproduced by CC calculations \cite{PhysRevLett.108.242501}.
The continuum shell model (CSM) using USDB interaction \cite{PhysRevLett.94.052501} provided a weakly unbound $^{26}$O ground state and unbound $^{28}$O.
Our results, obtained from GSM calculations, show the same trend as those arising from CSM for $^{24\text -28}$O.
Recently, \textit{ab initio} GSM calculations \cite{HU2020135206} based on realistic nuclear force, including  $pf$ partial waves in the GSM model space,
provided unbound $^{26,28}$O ground states by about 0.6 and 1.5 MeV with respect to the $^{24}$O ground state, respectively.

In Fig. \ref{Binding_O24}, GSM calculations using the $A$-dependent EFT(318) and the $A$-independent EFT(356), EFT(390), and EFT(436) interactions provide good agreements of the $^{23\text -26}$O ground states with experimental data. Furthermore, all the calculations give similar results.
In the following, we will employ the EFT(356) interaction to calculate the low-lying states of the neutron-rich oxygen isotopes and associated observables.
We also checked that the calculated one-body densities are not sensitive to the  EFT interactions used.

\begin{figure}[!htb]
\includegraphics[width=1.00\columnwidth]{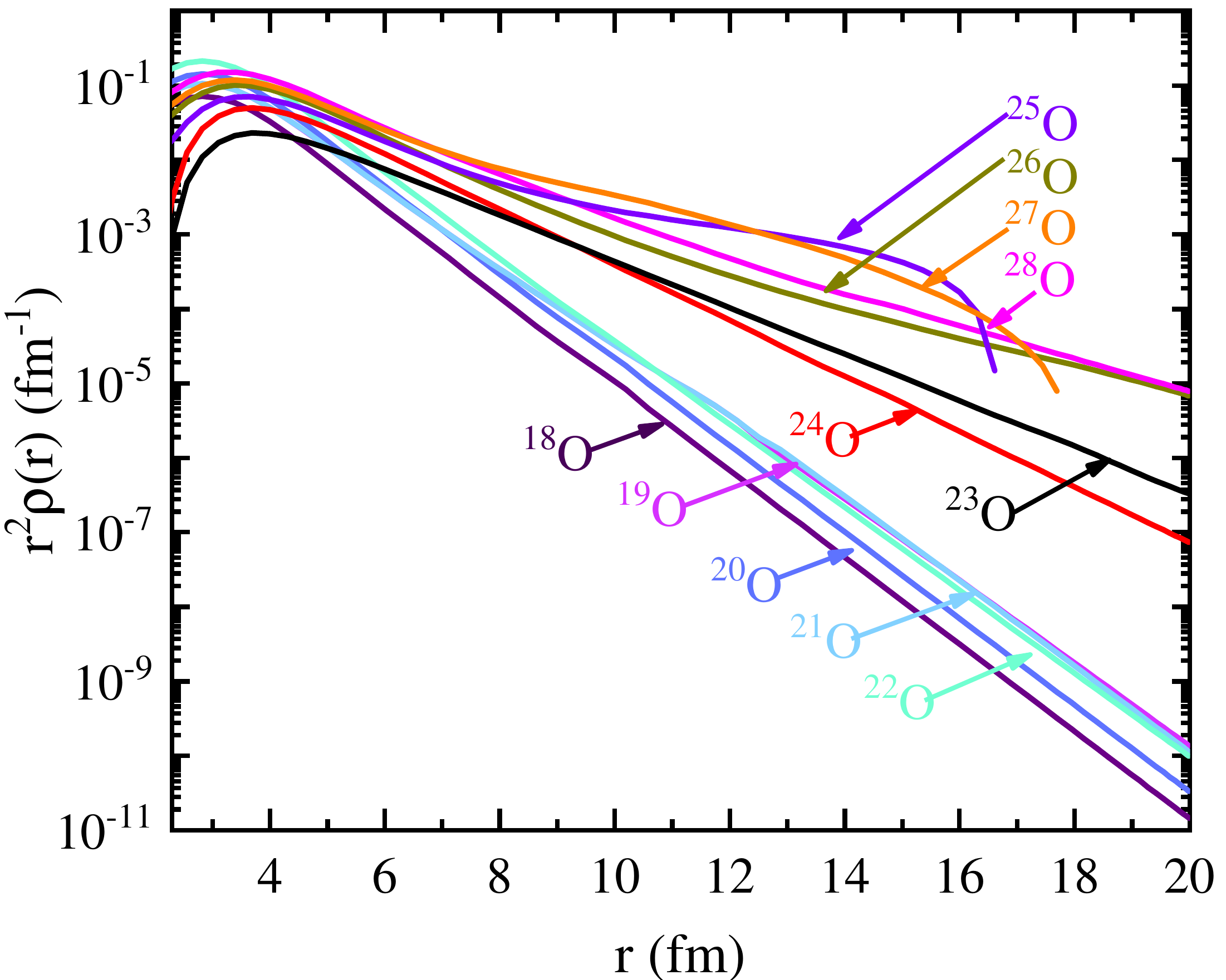}% Here is how to import EPS art
\caption{Radial one-body density multiplied by radius squared, denoted by $r^2\rho(r)$, of the ground states of oxygen neutron-rich $^{18\text- 28}$O isotopes as a function of the radius $r$ (in fm). }{\label{g.s_density_r2}}
\end{figure}

To further study dripline nuclei, we use GSM to calculate the one-body density of the ground states of the neutron-rich $^{18\text -28}$O oxygen isotopes.
The $^{18\text -22}$O isotopes are considered by using an $^{16}$O core and the same two-body EFT(356) interaction. Parameters of the WS potential of $^{16}$O are taken from Ref. \cite{PhysRevLett.89.042502}.
The calculated energies of the ground states of $^{18\text -22}$O have been checked to be close to experimental data.
The one-body densities of unbound states are complex in GSM, because unbound states are resonant and hence bear a complex energy.
The imaginary part of the calculated one-body density is small compared to the real part in a bound state or resonance with a narrow width. However the imaginary part of the one-body density of a resonance state with a large width is comparable with the real part in the asymptotic region, and the real part starts to oscillate in the asymptotic region.  We will only consider the real part of the one-body density in the following.
The results are shown in Fig. \ref{g.s_density_r2}.
The one-body densities of Fig. \ref{g.s_density_r2} support our first conclusion based on energetics only, i.e., that $^{25\text -28}$O are located beyond the neutron dripline and that $^{24}$O is the last bound oxygen isotope. Indeed, the one-body densities of the $^{18\text -22}$O are localized in the nuclear region (see Fig. \ref{g.s_density_r2}).
As continuum coupling is exactly treated, this localization property is not an artifact arising from basis-dependence, on the contrary, reflects the well bound, weakly bound or loosely unbound character of these nuclei.
It is clearly seen that the one-body densities of $^{25,27}$O, which start to oscillate in the asymptotic region, become negative after a given radius and can no longer be seen on the figure. These oscillations appear because $^{25,27}$O have  sizable resonance widths, close to 100 keV.

The ground states of $^{26}$O and $^{28}$O are unbound, but bear negligible widths.
Hence, their one-body densities resemble that of a loosely bound state, i.e., they decrease exponentially but slowly, so that they have a halo-like structure.
As one considers resonance states, these one-body densities will eventually start to increase in modulus without bounds.
However, due to the small character of particle-emission widths, this increase occurs for radius well beyond 20 fm.
Combining the results of Figs. \ref{Binding_O24} and \ref{g.s_density_r2}, we suggest that the ground state of $^{28}$O exhibits four-neutron decay by  way of $2n$-$2n$ emission via the $^{26}$O ground state. The four-neutron emission of the $^{28}$O ground state has also been suggested in other works, using few-body methods \cite{PhysRevC.84.021303} and  GSM using MBPT \cite{HU2020135206}.

\begin{figure}[!htb]
\includegraphics[width=1.00\columnwidth]{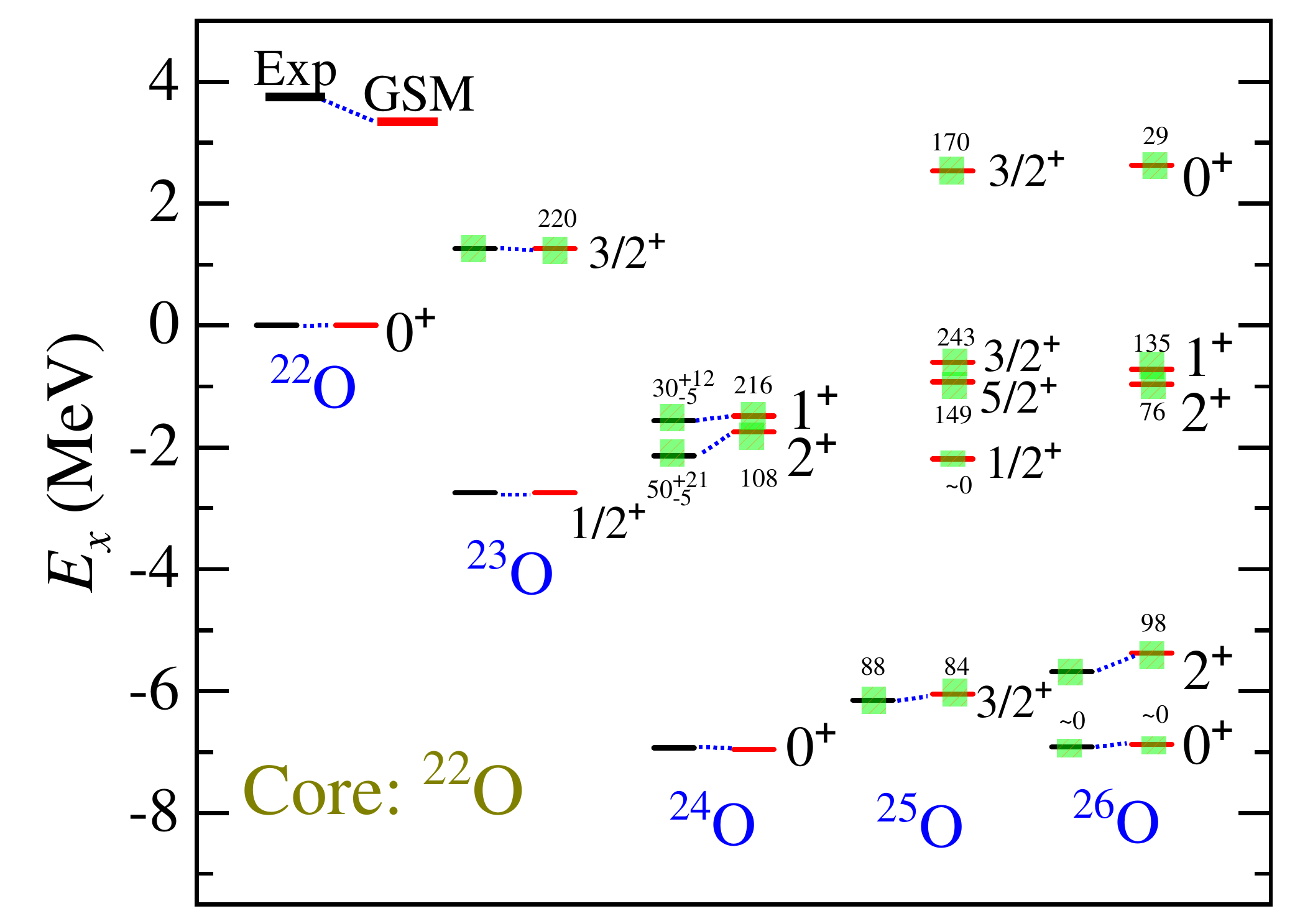}% Here is how to import EPS art
\caption{GSM calculations of low-lying positive-parity states of $^{23\text-26}$O using the EFT(356) interaction, along with experimental data \cite{PhysRevLett.116.102503,ensdf}.
 Resonance state is indicated with a green shade, and its width is written above (or below) the level in keV.}{\label{spectra}}
\end{figure}

Due to the good agreement with experimental data, obtained with GSM for binding energies of the neutron-rich oxygen isotopes using the  EFT(356) interaction,
one can make predictions concerning the unbound excited states of the $^{23\text -26}$O isotopes. Their spectra are presented in Fig. \ref{spectra}, along with available experimental data \cite{ensdf}.
For the $^{27,28}$O, only the ground states are considered in the present work, which are shown in the Fig. \ref{Binding_O24}.
Firstly, the experimentally known low-lying states of $^{23 \text -26}$O isotopes are well reproduced in GSM.
The largest theory-experiment differences  encountered in the  calculated energies are about  350 keV in $^{23\text -26}$O nuclei when compared with experimental data \cite{ensdf}.
One can give the example of the $2^+$ excited state of $^{24}$O (see Fig. \ref{spectra}).
Calculated widths are of the order of 10 to 100 keV, agreeing well with experimental data.
All the calculated unbound states have a width of at most 240 keV, while experimental widths do not exceed 100 keV.
Consequently, one can predict that the $^{23\text -26}$O isotopes possess  low-lying spectra consisting of narrow resonance states, whose widths should be expected to be about 50$-$100 keV.

\begin{figure}[!htb]
\includegraphics[width=1.00\columnwidth]{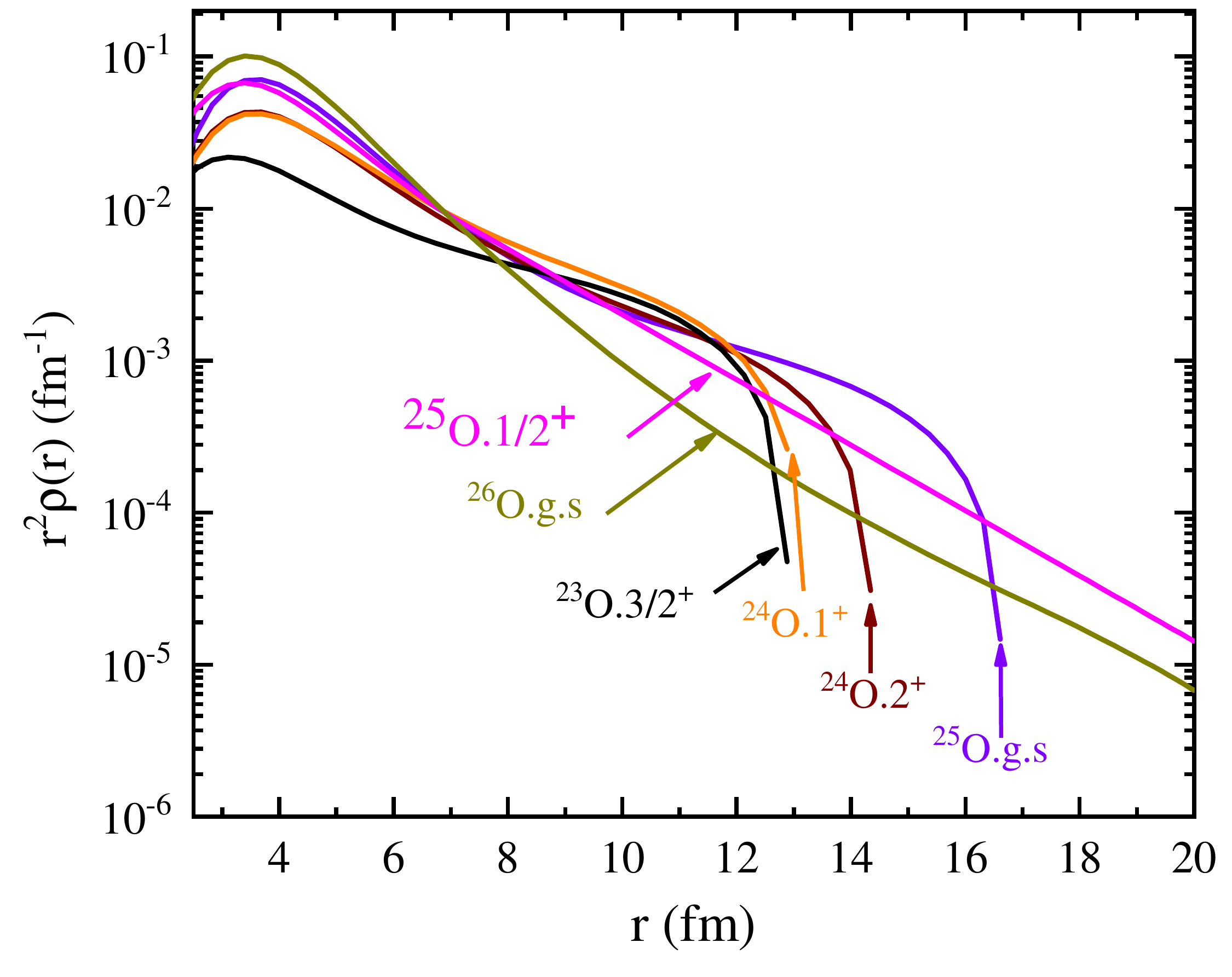}% Here is how to import EPS art
\caption{Similar to Fig. \ref{g.s_density_r2}, but for the excited states (and also the ground states in $^{25,26}$O for comparison) of the neutron-dripline oxygen isotopes calculated with GSM using the  EFT(356) interaction.}{\label{density_excited}}
\end{figure}

Similarly to the oxygen ground states of Fig. \ref{g.s_density_r2},
let us consider the one-body densities of the unbound spectra of the $^{23\text -26}$O isotopes (see Fig. \ref{density_excited}).
Due to the unbound character of excited states, their one-body densities typically oscillate in the asymptotic region.
One obtains the exponentially decaying densities in $r \leq 20$ fm for the excited $^{25}$O($1/2^+$) and the $^{26}$O ground states.
This is because their widths are almost zero.
The one-body densities which start to oscillate around $r \approx 13$ fm are associated to the eigenstates with large widths.
Indeed, the $^{23}$O($3/2^+$) and $^{24}$O($1^+$) excited state bear a width of about 230 keV, which is dominated by the occupation of the $0d_{3/2}$ orbital.
Comparatively, the  $^{25}$O ground state  and $^{24}$O($2^+$) excited state have widths equal to 64 keV and 87 keV, respectively, so that their densities start to oscillate at a slightly larger radius of $r \approx 16$ fm.
Consequently, the asymptotes of the one-body densities in $^{23\text
-26}$O isotopes are in accordance with the narrow or broad character of the calculated eigenstates.

An interesting excited state is  the ${1/2}^+$ excited state of $^{25}$O.
It is located above the ground state of $^{23}$O and below the excited states of $^{24}$O, with its main configuration being $^{22}$O$\otimes(1s_{1/2})^1(0d_{3/2})^2$. 
The $1s_{1/2}$ orbital is bound, and pairing makes the wave function of the two paired $0d_{3/2}$ neutrons less spreading in space. These lead to a reduction in the width of the ${1/2^+}$ excited configuration $^{22}$O$\otimes (1s_{1/2})^1(0d_{3/2})^2$.
The wave function of the ground state of $^{26}$O is dominated by the configuration $^{22}$O$\otimes (1s_{1/2})^2 (0d_{3/2})^2$, so that $^{25}$O($1/2^+$) and the ground state of $^{26}$O differ mainly by one well-bound neutron. Thus, due to a very small width of the $^{25}$O ${1/2}^+$ excited state and a large spectroscopic factor with the ground state of $^{26}$O, the ${1/2}^+$ excited state of $^{25}$O might be a candidate for two-neutron radioactivity at neutron dripline.

\section{Summary} \label{Summary}

We have applied GSM to  unbound neutron-rich oxygen isotopes, where continuum coupling and internucleon correlations are both included.
Due to the inert character of the ground state of $^{22}$O, we could build a Hamiltonian from a WS potential mimicking the effect of the $^{22}$O core,
above which valence neutrons interact. An EFT framework was used to generate interactions.
For the EFT interaction, only the leading-order  nucleon-nucleon interaction was considered.
Calculations with different interactions are performed.
It is found that an $A$-dependence in the EFT(318) two-body interaction is needed to compensate for the absence of three-body forces, while the dependence is not necessary when using the EFT(356), EFT(390), and EFT(436) interactions.

All the devised interactions can satisfactorily reproduce the energies of low-lying states in neutron-rich $^{23\text-26}$O isotopes,
EFT(356) interaction was used  to assess the asymptotic properties of the unbound ground and excited states of the $^{23\text-26}$O isotopes.
One can estimate the two-neutron separation energy of $^{26}$O, of the order of 18 keV.
$^{28}$O was found to be unbound by about 700 keV above the ground state of $^{24}$O.
Due to its very small particle-emission width, our calculations suggest that $^{28}$O has a four-neutron decay of the form $2n$-$2n$ via the ground state of $^{26}$O.
The $^{25\text{-}28}$O ground states are found to be resonances, so that one infers that $^{24}$O is the heaviest bound oxygen isotope.

It was also demonstrated that the asymptotes of one-body densities follow the narrow or broad resonance character of considered eigenstates.
Indeed, one-body densities either bear a slow exponential decay, similar to that appearing in halo nuclei, for moderate radii when the particle-emission width is negligible,
or exhibit oscillations after a radius $r \approx$ 10-15 fm in the case of resonances bearing a width larger than 50 keV typically.

We can make predictions of possible unbound excited states in $^{25,26}$O.
Indeed, one could identify low-lying excited states with widths smaller than 250 keV, which can be expected to be narrow resonances experimentally.
Moreover, the $^{25}$O(${1/2}^+$) excited state was found to bear a very small width and to differ from the $^{25}$O ground state mainly by a well bound $1s_{1/2}$ orbital.
Thus, the $^{25}$O(${1/2}^+$) excited state has a large spectroscopic factor with the ground state of $^{26}$O and then might be a two-neutron emitter.

GSM has showed to be very efficient to describe loosely bound and resonance states at neutron dripline, where both energies and widths can be well calculated.
The particle emission width obtained in the GSM is the total width which includes all possible partial decays.
It can be expected that the GSM calculations should be useful for future experiments dealing with particle emissions in nuclei around driplines.  \\

\textit{Acknowledgments} -- Valuable discussions with S. M. Wang, Q. Yuan, S. Zhang,  B. Dai, and Y. Z. Ma are gratefully acknowledged.
This work has been supported by the National Key R\&D Program of China under Grant No. 2018YFA0404401; the National Natural Science Foundation of China under Grants No. 11835001, 11921006, 12035001, and 11975282; the State Key Laboratory of Nuclear Physics and Technology, Peking University under Grant No. NPT2020ZZ01 and NPT2020KFY13; the Strategic Priority Research Program of Chinese Academy of Sciences under Grant No. XDB34000000; and the CUSTIPEN (China-U.S. Theory Institute for Physics with Exotic Nuclei) funded by the U.S. Department of Energy, Office of Science under Grant No. de-sc0009971. We acknowledge the High-Performance Computing Platform of Peking University for providing computational resources.

\bibliography{O_revision3}

\begin{thebibliography}{50}%
\makeatletter
\providecommand \@ifxundefined [1]{%
 \@ifx{#1\undefined}
}%
\providecommand \@ifnum [1]{%
 \ifnum #1\expandafter \@firstoftwo
 \else \expandafter \@secondoftwo
 \fi
}%
\providecommand \@ifx [1]{%
 \ifx #1\expandafter \@firstoftwo
 \else \expandafter \@secondoftwo
 \fi
}%
\providecommand \natexlab [1]{#1}%
\providecommand \enquote  [1]{``#1''}%
\providecommand \bibnamefont  [1]{#1}%
\providecommand \bibfnamefont [1]{#1}%
\providecommand \citenamefont [1]{#1}%
\providecommand \href@noop [0]{\@secondoftwo}%
\providecommand \href [0]{\begingroup \@sanitize@url \@href}%
\providecommand \@href[1]{\@@startlink{#1}\@@href}%
\providecommand \@@href[1]{\endgroup#1\@@endlink}%
\providecommand \@sanitize@url [0]{\catcode `\\12\catcode `\$12\catcode
  `\&12\catcode `\#12\catcode `\^12\catcode `\_12\catcode `\%12\relax}%
\providecommand \@@startlink[1]{}%
\providecommand \@@endlink[0]{}%
\providecommand \url  [0]{\begingroup\@sanitize@url \@url }%
\providecommand \@url [1]{\endgroup\@href {#1}{\urlprefix }}%
\providecommand \urlprefix  [0]{URL }%
\providecommand \Eprint [0]{\href }%
\providecommand \doibase [0]{http://dx.doi.org/}%
\providecommand \selectlanguage [0]{\@gobble}%
\providecommand \bibinfo  [0]{\@secondoftwo}%
\providecommand \bibfield  [0]{\@secondoftwo}%
\providecommand \translation [1]{[#1]}%
\providecommand \BibitemOpen [0]{}%
\providecommand \bibitemStop [0]{}%
\providecommand \bibitemNoStop [0]{.\EOS\space}%
\providecommand \EOS [0]{\spacefactor3000\relax}%
\providecommand \BibitemShut  [1]{\csname bibitem#1\endcsname}%
\let\auto@bib@innerbib\@empty
%</preamble>
\bibitem [{\citenamefont {Becheva}\ \emph {et~al.}(2006)\citenamefont
  {Becheva}, \citenamefont {Blumenfeld}, \citenamefont {Khan}, \citenamefont
  {Beaumel}, \citenamefont {Daugas}, \citenamefont {Delaunay}, \citenamefont
  {Demonchy}, \citenamefont {Drouart}, \citenamefont {Fallot}, \citenamefont
  {Gillibert}, \citenamefont {Giot}, \citenamefont {Grasso}, \citenamefont
  {Keeley}, \citenamefont {Kemper}, \citenamefont {Khoa}, \citenamefont
  {Lapoux}, \citenamefont {Lima}, \citenamefont {Musumarra}, \citenamefont
  {Nalpas}, \citenamefont {Pollacco}, \citenamefont {Roig}, \citenamefont
  {Roussel-Chomaz}, \citenamefont {Sauvestre}, \citenamefont {Scarpaci},
  \citenamefont {Skaza},\ and\ \citenamefont {Than}}]{PhysRevLett.96.012501}%
  \BibitemOpen
  \bibfield  {author} {\bibinfo {author} {\bibfnamefont {E.}~\bibnamefont
  {Becheva}}, \bibinfo {author} {\bibfnamefont {Y.}~\bibnamefont {Blumenfeld}},
  \bibinfo {author} {\bibfnamefont {E.}~\bibnamefont {Khan}}, \bibinfo {author}
  {\bibfnamefont {D.}~\bibnamefont {Beaumel}}, \bibinfo {author} {\bibfnamefont
  {J.~M.}\ \bibnamefont {Daugas}}, \bibinfo {author} {\bibfnamefont
  {F.}~\bibnamefont {Delaunay}}, \bibinfo {author} {\bibfnamefont {C.-E.}\
  \bibnamefont {Demonchy}}, \bibinfo {author} {\bibfnamefont {A.}~\bibnamefont
  {Drouart}}, \bibinfo {author} {\bibfnamefont {M.}~\bibnamefont {Fallot}},
  \bibinfo {author} {\bibfnamefont {A.}~\bibnamefont {Gillibert}}, \bibinfo
  {author} {\bibfnamefont {L.}~\bibnamefont {Giot}}, \bibinfo {author}
  {\bibfnamefont {M.}~\bibnamefont {Grasso}}, \bibinfo {author} {\bibfnamefont
  {N.}~\bibnamefont {Keeley}}, \bibinfo {author} {\bibfnamefont {K.~W.}\
  \bibnamefont {Kemper}}, \bibinfo {author} {\bibfnamefont {D.~T.}\
  \bibnamefont {Khoa}}, \bibinfo {author} {\bibfnamefont {V.}~\bibnamefont
  {Lapoux}}, \bibinfo {author} {\bibfnamefont {V.}~\bibnamefont {Lima}},
  \bibinfo {author} {\bibfnamefont {A.}~\bibnamefont {Musumarra}}, \bibinfo
  {author} {\bibfnamefont {L.}~\bibnamefont {Nalpas}}, \bibinfo {author}
  {\bibfnamefont {E.~C.}\ \bibnamefont {Pollacco}}, \bibinfo {author}
  {\bibfnamefont {O.}~\bibnamefont {Roig}}, \bibinfo {author} {\bibfnamefont
  {P.}~\bibnamefont {Roussel-Chomaz}}, \bibinfo {author} {\bibfnamefont
  {J.~E.}\ \bibnamefont {Sauvestre}}, \bibinfo {author} {\bibfnamefont {J.~A.}\
  \bibnamefont {Scarpaci}}, \bibinfo {author} {\bibfnamefont {F.}~\bibnamefont
  {Skaza}}, \ and\ \bibinfo {author} {\bibfnamefont {H.~S.}\ \bibnamefont
  {Than}},\ }\href {\doibase 10.1103/PhysRevLett.96.012501} {\bibfield
  {journal} {\bibinfo  {journal} {Phys. Rev. Lett.}\ }\textbf {\bibinfo
  {volume} {96}},\ \bibinfo {pages} {012501} (\bibinfo {year}
  {2006})}\BibitemShut {NoStop}%
\bibitem [{\citenamefont {Stanoiu}\ \emph {et~al.}(2004)\citenamefont
  {Stanoiu}, \citenamefont {Azaiez}, \citenamefont {Dombr\'adi}, \citenamefont
  {Sorlin}, \citenamefont {Brown}, \citenamefont {Belleguic}, \citenamefont
  {Sohler}, \citenamefont {Saint~Laurent}, \citenamefont {Lopez-Jimenez},
  \citenamefont {Penionzhkevich}, \citenamefont {Sletten}, \citenamefont
  {Achouri}, \citenamefont {Ang\'elique}, \citenamefont {Becker}, \citenamefont
  {Borcea}, \citenamefont {Bourgeois}, \citenamefont {Bracco}, \citenamefont
  {Daugas}, \citenamefont {Dlouh\'y}, \citenamefont {Donzaud}, \citenamefont
  {Duprat}, \citenamefont {F\"ul\"op}, \citenamefont {Guillemaud-Mueller},
  \citenamefont {Gr\'evy}, \citenamefont {Ibrahim}, \citenamefont {Kerek},
  \citenamefont {Krasznahorkay}, \citenamefont {Lewitowicz}, \citenamefont
  {Leenhardt}, \citenamefont {Lukyanov}, \citenamefont {Mayet}, \citenamefont
  {Mandal}, \citenamefont {van~der Marel}, \citenamefont {Mittig},
  \citenamefont {Mr\'azek}, \citenamefont {Negoita}, \citenamefont
  {De~Oliveira-Santos}, \citenamefont {Podoly\'ak}, \citenamefont {Pougheon},
  \citenamefont {Porquet}, \citenamefont {Roussel-Chomaz}, \citenamefont
  {Savajols}, \citenamefont {Sobolev}, \citenamefont {Stodel}, \citenamefont
  {Tim\'ar},\ and\ \citenamefont {Yamamoto}}]{PhysRevC.69.034312}%
  \BibitemOpen
  \bibfield  {author} {\bibinfo {author} {\bibfnamefont {M.}~\bibnamefont
  {Stanoiu}}, \bibinfo {author} {\bibfnamefont {F.}~\bibnamefont {Azaiez}},
  \bibinfo {author} {\bibfnamefont {Z.}~\bibnamefont {Dombr\'adi}}, \bibinfo
  {author} {\bibfnamefont {O.}~\bibnamefont {Sorlin}}, \bibinfo {author}
  {\bibfnamefont {B.~A.}\ \bibnamefont {Brown}}, \bibinfo {author}
  {\bibfnamefont {M.}~\bibnamefont {Belleguic}}, \bibinfo {author}
  {\bibfnamefont {D.}~\bibnamefont {Sohler}}, \bibinfo {author} {\bibfnamefont
  {M.~G.}\ \bibnamefont {Saint~Laurent}}, \bibinfo {author} {\bibfnamefont
  {M.~J.}\ \bibnamefont {Lopez-Jimenez}}, \bibinfo {author} {\bibfnamefont
  {Y.~E.}\ \bibnamefont {Penionzhkevich}}, \bibinfo {author} {\bibfnamefont
  {G.}~\bibnamefont {Sletten}}, \bibinfo {author} {\bibfnamefont {N.~L.}\
  \bibnamefont {Achouri}}, \bibinfo {author} {\bibfnamefont {J.~C.}\
  \bibnamefont {Ang\'elique}}, \bibinfo {author} {\bibfnamefont
  {F.}~\bibnamefont {Becker}}, \bibinfo {author} {\bibfnamefont
  {C.}~\bibnamefont {Borcea}}, \bibinfo {author} {\bibfnamefont
  {C.}~\bibnamefont {Bourgeois}}, \bibinfo {author} {\bibfnamefont
  {A.}~\bibnamefont {Bracco}}, \bibinfo {author} {\bibfnamefont {J.~M.}\
  \bibnamefont {Daugas}}, \bibinfo {author} {\bibfnamefont {Z.}~\bibnamefont
  {Dlouh\'y}}, \bibinfo {author} {\bibfnamefont {C.}~\bibnamefont {Donzaud}},
  \bibinfo {author} {\bibfnamefont {J.}~\bibnamefont {Duprat}}, \bibinfo
  {author} {\bibfnamefont {Z.}~\bibnamefont {F\"ul\"op}}, \bibinfo {author}
  {\bibfnamefont {D.}~\bibnamefont {Guillemaud-Mueller}}, \bibinfo {author}
  {\bibfnamefont {S.}~\bibnamefont {Gr\'evy}}, \bibinfo {author} {\bibfnamefont
  {F.}~\bibnamefont {Ibrahim}}, \bibinfo {author} {\bibfnamefont
  {A.}~\bibnamefont {Kerek}}, \bibinfo {author} {\bibfnamefont
  {A.}~\bibnamefont {Krasznahorkay}}, \bibinfo {author} {\bibfnamefont
  {M.}~\bibnamefont {Lewitowicz}}, \bibinfo {author} {\bibfnamefont
  {S.}~\bibnamefont {Leenhardt}}, \bibinfo {author} {\bibfnamefont
  {S.}~\bibnamefont {Lukyanov}}, \bibinfo {author} {\bibfnamefont
  {P.}~\bibnamefont {Mayet}}, \bibinfo {author} {\bibfnamefont
  {S.}~\bibnamefont {Mandal}}, \bibinfo {author} {\bibfnamefont
  {H.}~\bibnamefont {van~der Marel}}, \bibinfo {author} {\bibfnamefont
  {W.}~\bibnamefont {Mittig}}, \bibinfo {author} {\bibfnamefont
  {J.}~\bibnamefont {Mr\'azek}}, \bibinfo {author} {\bibfnamefont
  {F.}~\bibnamefont {Negoita}}, \bibinfo {author} {\bibfnamefont
  {F.}~\bibnamefont {De~Oliveira-Santos}}, \bibinfo {author} {\bibfnamefont
  {Z.}~\bibnamefont {Podoly\'ak}}, \bibinfo {author} {\bibfnamefont
  {F.}~\bibnamefont {Pougheon}}, \bibinfo {author} {\bibfnamefont {M.~G.}\
  \bibnamefont {Porquet}}, \bibinfo {author} {\bibfnamefont {P.}~\bibnamefont
  {Roussel-Chomaz}}, \bibinfo {author} {\bibfnamefont {H.}~\bibnamefont
  {Savajols}}, \bibinfo {author} {\bibfnamefont {Y.}~\bibnamefont {Sobolev}},
  \bibinfo {author} {\bibfnamefont {C.}~\bibnamefont {Stodel}}, \bibinfo
  {author} {\bibfnamefont {J.}~\bibnamefont {Tim\'ar}}, \ and\ \bibinfo
  {author} {\bibfnamefont {A.}~\bibnamefont {Yamamoto}},\ }\href {\doibase
  10.1103/PhysRevC.69.034312} {\bibfield  {journal} {\bibinfo  {journal} {Phys.
  Rev. C}\ }\textbf {\bibinfo {volume} {69}},\ \bibinfo {pages} {034312}
  (\bibinfo {year} {2004})}\BibitemShut {NoStop}%
\bibitem [{\citenamefont {Hoffman}\ \emph {et~al.}(2008)\citenamefont
  {Hoffman}, \citenamefont {Baumann}, \citenamefont {Bazin}, \citenamefont
  {Brown}, \citenamefont {Christian}, \citenamefont {DeYoung}, \citenamefont
  {Finck}, \citenamefont {Frank}, \citenamefont {Hinnefeld}, \citenamefont
  {Howes}, \citenamefont {Mears}, \citenamefont {Mosby}, \citenamefont {Mosby},
  \citenamefont {Reith}, \citenamefont {Rizzo}, \citenamefont {Rogers},
  \citenamefont {Peaslee}, \citenamefont {Peters}, \citenamefont {Schiller},
  \citenamefont {Scott}, \citenamefont {Tabor}, \citenamefont {Thoennessen},
  \citenamefont {Voss},\ and\ \citenamefont
  {Williams}}]{PhysRevLett.100.152502}%
  \BibitemOpen
  \bibfield  {author} {\bibinfo {author} {\bibfnamefont {C.~R.}\ \bibnamefont
  {Hoffman}}, \bibinfo {author} {\bibfnamefont {T.}~\bibnamefont {Baumann}},
  \bibinfo {author} {\bibfnamefont {D.}~\bibnamefont {Bazin}}, \bibinfo
  {author} {\bibfnamefont {J.}~\bibnamefont {Brown}}, \bibinfo {author}
  {\bibfnamefont {G.}~\bibnamefont {Christian}}, \bibinfo {author}
  {\bibfnamefont {P.~A.}\ \bibnamefont {DeYoung}}, \bibinfo {author}
  {\bibfnamefont {J.~E.}\ \bibnamefont {Finck}}, \bibinfo {author}
  {\bibfnamefont {N.}~\bibnamefont {Frank}}, \bibinfo {author} {\bibfnamefont
  {J.}~\bibnamefont {Hinnefeld}}, \bibinfo {author} {\bibfnamefont
  {R.}~\bibnamefont {Howes}}, \bibinfo {author} {\bibfnamefont
  {P.}~\bibnamefont {Mears}}, \bibinfo {author} {\bibfnamefont
  {E.}~\bibnamefont {Mosby}}, \bibinfo {author} {\bibfnamefont
  {S.}~\bibnamefont {Mosby}}, \bibinfo {author} {\bibfnamefont
  {J.}~\bibnamefont {Reith}}, \bibinfo {author} {\bibfnamefont
  {B.}~\bibnamefont {Rizzo}}, \bibinfo {author} {\bibfnamefont {W.~F.}\
  \bibnamefont {Rogers}}, \bibinfo {author} {\bibfnamefont {G.}~\bibnamefont
  {Peaslee}}, \bibinfo {author} {\bibfnamefont {W.~A.}\ \bibnamefont {Peters}},
  \bibinfo {author} {\bibfnamefont {A.}~\bibnamefont {Schiller}}, \bibinfo
  {author} {\bibfnamefont {M.~J.}\ \bibnamefont {Scott}}, \bibinfo {author}
  {\bibfnamefont {S.~L.}\ \bibnamefont {Tabor}}, \bibinfo {author}
  {\bibfnamefont {M.}~\bibnamefont {Thoennessen}}, \bibinfo {author}
  {\bibfnamefont {P.~J.}\ \bibnamefont {Voss}}, \ and\ \bibinfo {author}
  {\bibfnamefont {T.}~\bibnamefont {Williams}},\ }\href {\doibase
  10.1103/PhysRevLett.100.152502} {\bibfield  {journal} {\bibinfo  {journal}
  {Phys. Rev. Lett.}\ }\textbf {\bibinfo {volume} {100}},\ \bibinfo {pages}
  {152502} (\bibinfo {year} {2008})}\BibitemShut {NoStop}%
\bibitem [{\citenamefont {Tshoo}\ \emph {et~al.}(2012)\citenamefont {Tshoo},
  \citenamefont {Satou}, \citenamefont {Bhang}, \citenamefont {Choi},
  \citenamefont {Nakamura}, \citenamefont {Kondo}, \citenamefont {Deguchi},
  \citenamefont {Kawada}, \citenamefont {Kobayashi}, \citenamefont {Nakayama},
  \citenamefont {Tanaka}, \citenamefont {Tanaka}, \citenamefont {Aoi},
  \citenamefont {Ishihara}, \citenamefont {Motobayashi}, \citenamefont {Otsu},
  \citenamefont {Sakurai}, \citenamefont {Takeuchi}, \citenamefont {Togano},
  \citenamefont {Yoneda}, \citenamefont {Li}, \citenamefont {Delaunay},
  \citenamefont {Gibelin}, \citenamefont {Marqu\'es}, \citenamefont {Orr},
  \citenamefont {Honda}, \citenamefont {Matsushita}, \citenamefont {Kobayashi},
  \citenamefont {Miyashita}, \citenamefont {Sumikama}, \citenamefont
  {Yoshinaga}, \citenamefont {Shimoura}, \citenamefont {Sohler}, \citenamefont
  {Zheng},\ and\ \citenamefont {Cao}}]{PhysRevLett.109.022501}%
  \BibitemOpen
  \bibfield  {author} {\bibinfo {author} {\bibfnamefont {K.}~\bibnamefont
  {Tshoo}}, \bibinfo {author} {\bibfnamefont {Y.}~\bibnamefont {Satou}},
  \bibinfo {author} {\bibfnamefont {H.}~\bibnamefont {Bhang}}, \bibinfo
  {author} {\bibfnamefont {S.}~\bibnamefont {Choi}}, \bibinfo {author}
  {\bibfnamefont {T.}~\bibnamefont {Nakamura}}, \bibinfo {author}
  {\bibfnamefont {Y.}~\bibnamefont {Kondo}}, \bibinfo {author} {\bibfnamefont
  {S.}~\bibnamefont {Deguchi}}, \bibinfo {author} {\bibfnamefont
  {Y.}~\bibnamefont {Kawada}}, \bibinfo {author} {\bibfnamefont
  {N.}~\bibnamefont {Kobayashi}}, \bibinfo {author} {\bibfnamefont
  {Y.}~\bibnamefont {Nakayama}}, \bibinfo {author} {\bibfnamefont {K.~N.}\
  \bibnamefont {Tanaka}}, \bibinfo {author} {\bibfnamefont {N.}~\bibnamefont
  {Tanaka}}, \bibinfo {author} {\bibfnamefont {N.}~\bibnamefont {Aoi}},
  \bibinfo {author} {\bibfnamefont {M.}~\bibnamefont {Ishihara}}, \bibinfo
  {author} {\bibfnamefont {T.}~\bibnamefont {Motobayashi}}, \bibinfo {author}
  {\bibfnamefont {H.}~\bibnamefont {Otsu}}, \bibinfo {author} {\bibfnamefont
  {H.}~\bibnamefont {Sakurai}}, \bibinfo {author} {\bibfnamefont
  {S.}~\bibnamefont {Takeuchi}}, \bibinfo {author} {\bibfnamefont
  {Y.}~\bibnamefont {Togano}}, \bibinfo {author} {\bibfnamefont
  {K.}~\bibnamefont {Yoneda}}, \bibinfo {author} {\bibfnamefont {Z.~H.}\
  \bibnamefont {Li}}, \bibinfo {author} {\bibfnamefont {F.}~\bibnamefont
  {Delaunay}}, \bibinfo {author} {\bibfnamefont {J.}~\bibnamefont {Gibelin}},
  \bibinfo {author} {\bibfnamefont {F.~M.}\ \bibnamefont {Marqu\'es}}, \bibinfo
  {author} {\bibfnamefont {N.~A.}\ \bibnamefont {Orr}}, \bibinfo {author}
  {\bibfnamefont {T.}~\bibnamefont {Honda}}, \bibinfo {author} {\bibfnamefont
  {M.}~\bibnamefont {Matsushita}}, \bibinfo {author} {\bibfnamefont
  {T.}~\bibnamefont {Kobayashi}}, \bibinfo {author} {\bibfnamefont
  {Y.}~\bibnamefont {Miyashita}}, \bibinfo {author} {\bibfnamefont
  {T.}~\bibnamefont {Sumikama}}, \bibinfo {author} {\bibfnamefont
  {K.}~\bibnamefont {Yoshinaga}}, \bibinfo {author} {\bibfnamefont
  {S.}~\bibnamefont {Shimoura}}, \bibinfo {author} {\bibfnamefont
  {D.}~\bibnamefont {Sohler}}, \bibinfo {author} {\bibfnamefont
  {T.}~\bibnamefont {Zheng}}, \ and\ \bibinfo {author} {\bibfnamefont {Z.~X.}\
  \bibnamefont {Cao}},\ }\href {\doibase 10.1103/PhysRevLett.109.022501}
  {\bibfield  {journal} {\bibinfo  {journal} {Phys. Rev. Lett.}\ }\textbf
  {\bibinfo {volume} {109}},\ \bibinfo {pages} {022501} (\bibinfo {year}
  {2012})}\BibitemShut {NoStop}%
\bibitem [{\citenamefont {Kondo}\ \emph {et~al.}(2016)\citenamefont {Kondo},
  \citenamefont {Nakamura}, \citenamefont {Tanaka}, \citenamefont {Minakata},
  \citenamefont {Ogoshi}, \citenamefont {Orr}, \citenamefont {Achouri},
  \citenamefont {Aumann}, \citenamefont {Baba}, \citenamefont {Delaunay},
  \citenamefont {Doornenbal}, \citenamefont {Fukuda}, \citenamefont {Gibelin},
  \citenamefont {Hwang}, \citenamefont {Inabe}, \citenamefont {Isobe},
  \citenamefont {Kameda}, \citenamefont {Kanno}, \citenamefont {Kim},
  \citenamefont {Kobayashi}, \citenamefont {Kobayashi}, \citenamefont {Kubo},
  \citenamefont {Leblond}, \citenamefont {Lee}, \citenamefont {Marqu\'es},
  \citenamefont {Motobayashi}, \citenamefont {Murai}, \citenamefont {Murakami},
  \citenamefont {Muto}, \citenamefont {Nakashima}, \citenamefont {Nakatsuka},
  \citenamefont {Navin}, \citenamefont {Nishi}, \citenamefont {Otsu},
  \citenamefont {Sato}, \citenamefont {Satou}, \citenamefont {Shimizu},
  \citenamefont {Suzuki}, \citenamefont {Takahashi}, \citenamefont {Takeda},
  \citenamefont {Takeuchi}, \citenamefont {Togano}, \citenamefont {Tuff},
  \citenamefont {Vandebrouck},\ and\ \citenamefont
  {Yoneda}}]{PhysRevLett.116.102503}%
  \BibitemOpen
  \bibfield  {author} {\bibinfo {author} {\bibfnamefont {Y.}~\bibnamefont
  {Kondo}}, \bibinfo {author} {\bibfnamefont {T.}~\bibnamefont {Nakamura}},
  \bibinfo {author} {\bibfnamefont {R.}~\bibnamefont {Tanaka}}, \bibinfo
  {author} {\bibfnamefont {R.}~\bibnamefont {Minakata}}, \bibinfo {author}
  {\bibfnamefont {S.}~\bibnamefont {Ogoshi}}, \bibinfo {author} {\bibfnamefont
  {N.~A.}\ \bibnamefont {Orr}}, \bibinfo {author} {\bibfnamefont {N.~L.}\
  \bibnamefont {Achouri}}, \bibinfo {author} {\bibfnamefont {T.}~\bibnamefont
  {Aumann}}, \bibinfo {author} {\bibfnamefont {H.}~\bibnamefont {Baba}},
  \bibinfo {author} {\bibfnamefont {F.}~\bibnamefont {Delaunay}}, \bibinfo
  {author} {\bibfnamefont {P.}~\bibnamefont {Doornenbal}}, \bibinfo {author}
  {\bibfnamefont {N.}~\bibnamefont {Fukuda}}, \bibinfo {author} {\bibfnamefont
  {J.}~\bibnamefont {Gibelin}}, \bibinfo {author} {\bibfnamefont {J.~W.}\
  \bibnamefont {Hwang}}, \bibinfo {author} {\bibfnamefont {N.}~\bibnamefont
  {Inabe}}, \bibinfo {author} {\bibfnamefont {T.}~\bibnamefont {Isobe}},
  \bibinfo {author} {\bibfnamefont {D.}~\bibnamefont {Kameda}}, \bibinfo
  {author} {\bibfnamefont {D.}~\bibnamefont {Kanno}}, \bibinfo {author}
  {\bibfnamefont {S.}~\bibnamefont {Kim}}, \bibinfo {author} {\bibfnamefont
  {N.}~\bibnamefont {Kobayashi}}, \bibinfo {author} {\bibfnamefont
  {T.}~\bibnamefont {Kobayashi}}, \bibinfo {author} {\bibfnamefont
  {T.}~\bibnamefont {Kubo}}, \bibinfo {author} {\bibfnamefont {S.}~\bibnamefont
  {Leblond}}, \bibinfo {author} {\bibfnamefont {J.}~\bibnamefont {Lee}},
  \bibinfo {author} {\bibfnamefont {F.~M.}\ \bibnamefont {Marqu\'es}}, \bibinfo
  {author} {\bibfnamefont {T.}~\bibnamefont {Motobayashi}}, \bibinfo {author}
  {\bibfnamefont {D.}~\bibnamefont {Murai}}, \bibinfo {author} {\bibfnamefont
  {T.}~\bibnamefont {Murakami}}, \bibinfo {author} {\bibfnamefont
  {K.}~\bibnamefont {Muto}}, \bibinfo {author} {\bibfnamefont {T.}~\bibnamefont
  {Nakashima}}, \bibinfo {author} {\bibfnamefont {N.}~\bibnamefont
  {Nakatsuka}}, \bibinfo {author} {\bibfnamefont {A.}~\bibnamefont {Navin}},
  \bibinfo {author} {\bibfnamefont {S.}~\bibnamefont {Nishi}}, \bibinfo
  {author} {\bibfnamefont {H.}~\bibnamefont {Otsu}}, \bibinfo {author}
  {\bibfnamefont {H.}~\bibnamefont {Sato}}, \bibinfo {author} {\bibfnamefont
  {Y.}~\bibnamefont {Satou}}, \bibinfo {author} {\bibfnamefont
  {Y.}~\bibnamefont {Shimizu}}, \bibinfo {author} {\bibfnamefont
  {H.}~\bibnamefont {Suzuki}}, \bibinfo {author} {\bibfnamefont
  {K.}~\bibnamefont {Takahashi}}, \bibinfo {author} {\bibfnamefont
  {H.}~\bibnamefont {Takeda}}, \bibinfo {author} {\bibfnamefont
  {S.}~\bibnamefont {Takeuchi}}, \bibinfo {author} {\bibfnamefont
  {Y.}~\bibnamefont {Togano}}, \bibinfo {author} {\bibfnamefont {A.~G.}\
  \bibnamefont {Tuff}}, \bibinfo {author} {\bibfnamefont {M.}~\bibnamefont
  {Vandebrouck}}, \ and\ \bibinfo {author} {\bibfnamefont {K.}~\bibnamefont
  {Yoneda}},\ }\href {\doibase 10.1103/PhysRevLett.116.102503} {\bibfield
  {journal} {\bibinfo  {journal} {Phys. Rev. Lett.}\ }\textbf {\bibinfo
  {volume} {116}},\ \bibinfo {pages} {102503} (\bibinfo {year}
  {2016})}\BibitemShut {NoStop}%
\bibitem [{\citenamefont {Kohley}\ \emph {et~al.}(2013)\citenamefont {Kohley},
  \citenamefont {Baumann}, \citenamefont {Bazin}, \citenamefont {Christian},
  \citenamefont {DeYoung}, \citenamefont {Finck}, \citenamefont {Frank},
  \citenamefont {Jones}, \citenamefont {Lunderberg}, \citenamefont {Luther},
  \citenamefont {Mosby}, \citenamefont {Nagi}, \citenamefont {Smith},
  \citenamefont {Snyder}, \citenamefont {Spyrou},\ and\ \citenamefont
  {Thoennessen}}]{PhysRevLett.110.152501}%
  \BibitemOpen
  \bibfield  {author} {\bibinfo {author} {\bibfnamefont {Z.}~\bibnamefont
  {Kohley}}, \bibinfo {author} {\bibfnamefont {T.}~\bibnamefont {Baumann}},
  \bibinfo {author} {\bibfnamefont {D.}~\bibnamefont {Bazin}}, \bibinfo
  {author} {\bibfnamefont {G.}~\bibnamefont {Christian}}, \bibinfo {author}
  {\bibfnamefont {P.~A.}\ \bibnamefont {DeYoung}}, \bibinfo {author}
  {\bibfnamefont {J.~E.}\ \bibnamefont {Finck}}, \bibinfo {author}
  {\bibfnamefont {N.}~\bibnamefont {Frank}}, \bibinfo {author} {\bibfnamefont
  {M.}~\bibnamefont {Jones}}, \bibinfo {author} {\bibfnamefont
  {E.}~\bibnamefont {Lunderberg}}, \bibinfo {author} {\bibfnamefont
  {B.}~\bibnamefont {Luther}}, \bibinfo {author} {\bibfnamefont
  {S.}~\bibnamefont {Mosby}}, \bibinfo {author} {\bibfnamefont
  {T.}~\bibnamefont {Nagi}}, \bibinfo {author} {\bibfnamefont {J.~K.}\
  \bibnamefont {Smith}}, \bibinfo {author} {\bibfnamefont {J.}~\bibnamefont
  {Snyder}}, \bibinfo {author} {\bibfnamefont {A.}~\bibnamefont {Spyrou}}, \
  and\ \bibinfo {author} {\bibfnamefont {M.}~\bibnamefont {Thoennessen}},\
  }\href {\doibase 10.1103/PhysRevLett.110.152501} {\bibfield  {journal}
  {\bibinfo  {journal} {Phys. Rev. Lett.}\ }\textbf {\bibinfo {volume} {110}},\
  \bibinfo {pages} {152501} (\bibinfo {year} {2013})}\BibitemShut {NoStop}%
\bibitem [{\citenamefont {Jones}\ \emph {et~al.}(2017)\citenamefont {Jones},
  \citenamefont {Fossez}, \citenamefont {Baumann}, \citenamefont {DeYoung},
  \citenamefont {Finck}, \citenamefont {Frank}, \citenamefont {Kuchera},
  \citenamefont {Michel}, \citenamefont {Nazarewicz}, \citenamefont {Rotureau},
  \citenamefont {Smith}, \citenamefont {Stephenson}, \citenamefont {Stiefel},
  \citenamefont {Thoennessen},\ and\ \citenamefont
  {Zegers}}]{PhysRevC.96.054322}%
  \BibitemOpen
  \bibfield  {author} {\bibinfo {author} {\bibfnamefont {M.~D.}\ \bibnamefont
  {Jones}}, \bibinfo {author} {\bibfnamefont {K.}~\bibnamefont {Fossez}},
  \bibinfo {author} {\bibfnamefont {T.}~\bibnamefont {Baumann}}, \bibinfo
  {author} {\bibfnamefont {P.~A.}\ \bibnamefont {DeYoung}}, \bibinfo {author}
  {\bibfnamefont {J.~E.}\ \bibnamefont {Finck}}, \bibinfo {author}
  {\bibfnamefont {N.}~\bibnamefont {Frank}}, \bibinfo {author} {\bibfnamefont
  {A.~N.}\ \bibnamefont {Kuchera}}, \bibinfo {author} {\bibfnamefont
  {N.}~\bibnamefont {Michel}}, \bibinfo {author} {\bibfnamefont
  {W.}~\bibnamefont {Nazarewicz}}, \bibinfo {author} {\bibfnamefont
  {J.}~\bibnamefont {Rotureau}}, \bibinfo {author} {\bibfnamefont {J.~K.}\
  \bibnamefont {Smith}}, \bibinfo {author} {\bibfnamefont {S.~L.}\ \bibnamefont
  {Stephenson}}, \bibinfo {author} {\bibfnamefont {K.}~\bibnamefont {Stiefel}},
  \bibinfo {author} {\bibfnamefont {M.}~\bibnamefont {Thoennessen}}, \ and\
  \bibinfo {author} {\bibfnamefont {R.~G.~T.}\ \bibnamefont {Zegers}},\ }\href
  {\doibase 10.1103/PhysRevC.96.054322} {\bibfield  {journal} {\bibinfo
  {journal} {Phys. Rev. C}\ }\textbf {\bibinfo {volume} {96}},\ \bibinfo
  {pages} {054322} (\bibinfo {year} {2017})}\BibitemShut {NoStop}%
\bibitem [{\citenamefont {Id~Betan}\ \emph {et~al.}(2002)\citenamefont
  {Id~Betan}, \citenamefont {Liotta}, \citenamefont {Sandulescu},\ and\
  \citenamefont {Vertse}}]{PhysRevLett.89.042501}%
  \BibitemOpen
  \bibfield  {author} {\bibinfo {author} {\bibfnamefont {R.}~\bibnamefont
  {Id~Betan}}, \bibinfo {author} {\bibfnamefont {R.~J.}\ \bibnamefont
  {Liotta}}, \bibinfo {author} {\bibfnamefont {N.}~\bibnamefont {Sandulescu}},
  \ and\ \bibinfo {author} {\bibfnamefont {T.}~\bibnamefont {Vertse}},\ }\href
  {\doibase 10.1103/PhysRevLett.89.042501} {\bibfield  {journal} {\bibinfo
  {journal} {Phys. Rev. Lett.}\ }\textbf {\bibinfo {volume} {89}},\ \bibinfo
  {pages} {042501} (\bibinfo {year} {2002})}\BibitemShut {NoStop}%
\bibitem [{\citenamefont {Michel}\ \emph {et~al.}(2002)\citenamefont {Michel},
  \citenamefont {Nazarewicz}, \citenamefont {P\l{}oszajczak},\ and\
  \citenamefont {Bennaceur}}]{PhysRevLett.89.042502}%
  \BibitemOpen
  \bibfield  {author} {\bibinfo {author} {\bibfnamefont {N.}~\bibnamefont
  {Michel}}, \bibinfo {author} {\bibfnamefont {W.}~\bibnamefont {Nazarewicz}},
  \bibinfo {author} {\bibfnamefont {M.}~\bibnamefont {P\l{}oszajczak}}, \ and\
  \bibinfo {author} {\bibfnamefont {K.}~\bibnamefont {Bennaceur}},\ }\href
  {\doibase 10.1103/PhysRevLett.89.042502} {\bibfield  {journal} {\bibinfo
  {journal} {Phys. Rev. Lett.}\ }\textbf {\bibinfo {volume} {89}},\ \bibinfo
  {pages} {042502} (\bibinfo {year} {2002})}\BibitemShut {NoStop}%
\bibitem [{\citenamefont {Forss\'en}\ \emph {et~al.}(2013)\citenamefont
  {Forss\'en}, \citenamefont {Hagen}, \citenamefont {Hjorth-Jensen},
  \citenamefont {Nazarewicz},\ and\ \citenamefont
  {Rotureau}}]{1402-4896-2013-T152-014022}%
  \BibitemOpen
  \bibfield  {author} {\bibinfo {author} {\bibfnamefont {C.}~\bibnamefont
  {Forss\'en}}, \bibinfo {author} {\bibfnamefont {G.}~\bibnamefont {Hagen}},
  \bibinfo {author} {\bibfnamefont {M.}~\bibnamefont {Hjorth-Jensen}}, \bibinfo
  {author} {\bibfnamefont {W.}~\bibnamefont {Nazarewicz}}, \ and\ \bibinfo
  {author} {\bibfnamefont {J.}~\bibnamefont {Rotureau}},\ }\href
  {http://stacks.iop.org/1402-4896/2013/i=T152/a=014022} {\bibfield  {journal}
  {\bibinfo  {journal} {Phys. Scri}\ }\textbf {\bibinfo {volume} {2013}},\
  \bibinfo {pages} {014022} (\bibinfo {year} {2013})}\BibitemShut {NoStop}%
\bibitem [{\citenamefont {Michel}\ \emph {et~al.}(2009)\citenamefont {Michel},
  \citenamefont {Nazarewicz}, \citenamefont {Płoszajczak},\ and\ \citenamefont
  {Vertse}}]{0954-3899-36-1-013101}%
  \BibitemOpen
  \bibfield  {author} {\bibinfo {author} {\bibfnamefont {N.}~\bibnamefont
  {Michel}}, \bibinfo {author} {\bibfnamefont {W.}~\bibnamefont {Nazarewicz}},
  \bibinfo {author} {\bibfnamefont {M.}~\bibnamefont {Płoszajczak}}, \ and\
  \bibinfo {author} {\bibfnamefont {T.}~\bibnamefont {Vertse}},\ }\href
  {http://stacks.iop.org/0954-3899/36/i=1/a=013101} {\bibfield  {journal}
  {\bibinfo  {journal} {Jour. Phys. G. Nucl. Part. Phys.}\ }\textbf {\bibinfo
  {volume} {36}},\ \bibinfo {pages} {013101} (\bibinfo {year}
  {2009})}\BibitemShut {NoStop}%
\bibitem [{\citenamefont {Berggren}(1968)}]{BERGGREN1968265}%
  \BibitemOpen
  \bibfield  {author} {\bibinfo {author} {\bibfnamefont {T.}~\bibnamefont
  {Berggren}},\ }\href {\doibase https://doi.org/10.1016/0375-9474(68)90593-9}
  {\bibfield  {journal} {\bibinfo  {journal} {Nucl. Phys. A}\ }\textbf
  {\bibinfo {volume} {109}},\ \bibinfo {pages} {265 } (\bibinfo {year}
  {1968})}\BibitemShut {NoStop}%
\bibitem [{\citenamefont {Papadimitriou}\ \emph {et~al.}(2011)\citenamefont
  {Papadimitriou}, \citenamefont {Kruppa}, \citenamefont {Michel},
  \citenamefont {Nazarewicz}, \citenamefont {P\l{}oszajczak},\ and\
  \citenamefont {Rotureau}}]{PhysRevC.84.051304}%
  \BibitemOpen
  \bibfield  {author} {\bibinfo {author} {\bibfnamefont {G.}~\bibnamefont
  {Papadimitriou}}, \bibinfo {author} {\bibfnamefont {A.~T.}\ \bibnamefont
  {Kruppa}}, \bibinfo {author} {\bibfnamefont {N.}~\bibnamefont {Michel}},
  \bibinfo {author} {\bibfnamefont {W.}~\bibnamefont {Nazarewicz}}, \bibinfo
  {author} {\bibfnamefont {M.}~\bibnamefont {P\l{}oszajczak}}, \ and\ \bibinfo
  {author} {\bibfnamefont {J.}~\bibnamefont {Rotureau}},\ }\href {\doibase
  10.1103/PhysRevC.84.051304} {\bibfield  {journal} {\bibinfo  {journal} {Phys.
  Rev. C}\ }\textbf {\bibinfo {volume} {84}},\ \bibinfo {pages} {051304(R)}
  (\bibinfo {year} {2011})}\BibitemShut {NoStop}%
\bibitem [{\citenamefont {Michel}\ \emph {et~al.}(2006)\citenamefont {Michel},
  \citenamefont {Nazarewicz}, \citenamefont {P\l{}oszajczak},\ and\
  \citenamefont {Rotureau}}]{PhysRevC.74.054305}%
  \BibitemOpen
  \bibfield  {author} {\bibinfo {author} {\bibfnamefont {N.}~\bibnamefont
  {Michel}}, \bibinfo {author} {\bibfnamefont {W.}~\bibnamefont {Nazarewicz}},
  \bibinfo {author} {\bibfnamefont {M.}~\bibnamefont {P\l{}oszajczak}}, \ and\
  \bibinfo {author} {\bibfnamefont {J.}~\bibnamefont {Rotureau}},\ }\href
  {\doibase 10.1103/PhysRevC.74.054305} {\bibfield  {journal} {\bibinfo
  {journal} {Phys. Rev. C}\ }\textbf {\bibinfo {volume} {74}},\ \bibinfo
  {pages} {054305} (\bibinfo {year} {2006})}\BibitemShut {NoStop}%
\bibitem [{\citenamefont {Jaganathen}\ \emph {et~al.}(2017)\citenamefont
  {Jaganathen}, \citenamefont {Id~Betan}, \citenamefont {Michel}, \citenamefont
  {Nazarewicz},\ and\ \citenamefont {P\l{}oszajczak}}]{PhysRevC.96.054316}%
  \BibitemOpen
  \bibfield  {author} {\bibinfo {author} {\bibfnamefont {Y.}~\bibnamefont
  {Jaganathen}}, \bibinfo {author} {\bibfnamefont {R.}~\bibnamefont
  {Id~Betan}}, \bibinfo {author} {\bibfnamefont {N.}~\bibnamefont {Michel}},
  \bibinfo {author} {\bibfnamefont {W.}~\bibnamefont {Nazarewicz}}, \ and\
  \bibinfo {author} {\bibfnamefont {M.}~\bibnamefont {P\l{}oszajczak}},\ }\href
  {\doibase 10.1103/PhysRevC.96.054316} {\bibfield  {journal} {\bibinfo
  {journal} {Phys. Rev. C}\ }\textbf {\bibinfo {volume} {96}},\ \bibinfo
  {pages} {054316} (\bibinfo {year} {2017})}\BibitemShut {NoStop}%
\bibitem [{\citenamefont {Sun}\ \emph {et~al.}(2017)\citenamefont {Sun},
  \citenamefont {Wu}, \citenamefont {Zhao}, \citenamefont {Hu}, \citenamefont
  {Dai},\ and\ \citenamefont {Xu}}]{SUN2017227}%
  \BibitemOpen
  \bibfield  {author} {\bibinfo {author} {\bibfnamefont {Z.~H.}\ \bibnamefont
  {Sun}}, \bibinfo {author} {\bibfnamefont {Q.}~\bibnamefont {Wu}}, \bibinfo
  {author} {\bibfnamefont {Z.~H.}\ \bibnamefont {Zhao}}, \bibinfo {author}
  {\bibfnamefont {B.~S.}\ \bibnamefont {Hu}}, \bibinfo {author} {\bibfnamefont
  {S.~J.}\ \bibnamefont {Dai}}, \ and\ \bibinfo {author} {\bibfnamefont
  {F.~R.}\ \bibnamefont {Xu}},\ }\href {\doibase
  https://doi.org/10.1016/j.physletb.2017.03.054} {\bibfield  {journal}
  {\bibinfo  {journal} {Phys. Lett. B}\ }\textbf {\bibinfo {volume} {769}},\
  \bibinfo {pages} {227 } (\bibinfo {year} {2017})}\BibitemShut {NoStop}%
\bibitem [{\citenamefont {Fossez}\ \emph {et~al.}(2017)\citenamefont {Fossez},
  \citenamefont {Rotureau}, \citenamefont {Michel},\ and\ \citenamefont
  {Nazarewicz}}]{PhysRevC.96.024308}%
  \BibitemOpen
  \bibfield  {author} {\bibinfo {author} {\bibfnamefont {K.}~\bibnamefont
  {Fossez}}, \bibinfo {author} {\bibfnamefont {J.}~\bibnamefont {Rotureau}},
  \bibinfo {author} {\bibfnamefont {N.}~\bibnamefont {Michel}}, \ and\ \bibinfo
  {author} {\bibfnamefont {W.}~\bibnamefont {Nazarewicz}},\ }\href {\doibase
  10.1103/PhysRevC.96.024308} {\bibfield  {journal} {\bibinfo  {journal} {Phys.
  Rev. C}\ }\textbf {\bibinfo {volume} {96}},\ \bibinfo {pages} {024308}
  (\bibinfo {year} {2017})}\BibitemShut {NoStop}%
\bibitem [{\citenamefont {Michel}\ \emph {et~al.}(2019)\citenamefont {Michel},
  \citenamefont {Li}, \citenamefont {Xu},\ and\ \citenamefont
  {Zuo}}]{PhysRevC.100.064303}%
  \BibitemOpen
  \bibfield  {author} {\bibinfo {author} {\bibfnamefont {N.}~\bibnamefont
  {Michel}}, \bibinfo {author} {\bibfnamefont {J.~G.}\ \bibnamefont {Li}},
  \bibinfo {author} {\bibfnamefont {F.~R.}\ \bibnamefont {Xu}}, \ and\ \bibinfo
  {author} {\bibfnamefont {W.}~\bibnamefont {Zuo}},\ }\href {\doibase
  10.1103/PhysRevC.100.064303} {\bibfield  {journal} {\bibinfo  {journal}
  {Phys. Rev. C}\ }\textbf {\bibinfo {volume} {100}},\ \bibinfo {pages}
  {064303} (\bibinfo {year} {2019})}\BibitemShut {NoStop}%
\bibitem [{\citenamefont {Machleidt}\ and\ \citenamefont
  {Entem}(2011)}]{MACHLEIDT20111}%
  \BibitemOpen
  \bibfield  {author} {\bibinfo {author} {\bibfnamefont {R.}~\bibnamefont
  {Machleidt}}\ and\ \bibinfo {author} {\bibfnamefont {D.}~\bibnamefont
  {Entem}},\ }\href {\doibase https://doi.org/10.1016/j.physrep.2011.02.001}
  {\bibfield  {journal} {\bibinfo  {journal} {Phys. Rep}\ }\textbf {\bibinfo
  {volume} {503}},\ \bibinfo {pages} {1 } (\bibinfo {year} {2011})}\BibitemShut
  {NoStop}%
\bibitem [{\citenamefont {Huth}\ \emph {et~al.}(2018)\citenamefont {Huth},
  \citenamefont {Durant}, \citenamefont {Simonis},\ and\ \citenamefont
  {Schwenk}}]{PhysRevC.98.044301}%
  \BibitemOpen
  \bibfield  {author} {\bibinfo {author} {\bibfnamefont {L.}~\bibnamefont
  {Huth}}, \bibinfo {author} {\bibfnamefont {V.}~\bibnamefont {Durant}},
  \bibinfo {author} {\bibfnamefont {J.}~\bibnamefont {Simonis}}, \ and\
  \bibinfo {author} {\bibfnamefont {A.}~\bibnamefont {Schwenk}},\ }\href
  {\doibase 10.1103/PhysRevC.98.044301} {\bibfield  {journal} {\bibinfo
  {journal} {Phys. Rev. C}\ }\textbf {\bibinfo {volume} {98}},\ \bibinfo
  {pages} {044301} (\bibinfo {year} {2018})}\BibitemShut {NoStop}%
\bibitem [{\citenamefont {Hagen}\ \emph {et~al.}(2012)\citenamefont {Hagen},
  \citenamefont {Hjorth-Jensen}, \citenamefont {Jansen}, \citenamefont
  {Machleidt},\ and\ \citenamefont {Papenbrock}}]{PhysRevLett.108.242501}%
  \BibitemOpen
  \bibfield  {author} {\bibinfo {author} {\bibfnamefont {G.}~\bibnamefont
  {Hagen}}, \bibinfo {author} {\bibfnamefont {M.}~\bibnamefont
  {Hjorth-Jensen}}, \bibinfo {author} {\bibfnamefont {G.~R.}\ \bibnamefont
  {Jansen}}, \bibinfo {author} {\bibfnamefont {R.}~\bibnamefont {Machleidt}}, \
  and\ \bibinfo {author} {\bibfnamefont {T.}~\bibnamefont {Papenbrock}},\
  }\href {\doibase 10.1103/PhysRevLett.108.242501} {\bibfield  {journal}
  {\bibinfo  {journal} {Phys. Rev. Lett.}\ }\textbf {\bibinfo {volume} {108}},\
  \bibinfo {pages} {242501} (\bibinfo {year} {2012})}\BibitemShut {NoStop}%
\bibitem [{\citenamefont {Tsukiyama}\ \emph {et~al.}(2009)\citenamefont
  {Tsukiyama}, \citenamefont {Hjorth-Jensen},\ and\ \citenamefont
  {Hagen}}]{PhysRevC.80.051301}%
  \BibitemOpen
  \bibfield  {author} {\bibinfo {author} {\bibfnamefont {K.}~\bibnamefont
  {Tsukiyama}}, \bibinfo {author} {\bibfnamefont {M.}~\bibnamefont
  {Hjorth-Jensen}}, \ and\ \bibinfo {author} {\bibfnamefont {G.}~\bibnamefont
  {Hagen}},\ }\href {\doibase 10.1103/PhysRevC.80.051301} {\bibfield  {journal}
  {\bibinfo  {journal} {Phys. Rev. C}\ }\textbf {\bibinfo {volume} {80}},\
  \bibinfo {pages} {051301(R)} (\bibinfo {year} {2009})}\BibitemShut {NoStop}%
\bibitem [{\citenamefont {Otsuka}\ \emph {et~al.}(2010)\citenamefont {Otsuka},
  \citenamefont {Suzuki}, \citenamefont {Holt}, \citenamefont {Schwenk},\ and\
  \citenamefont {Akaishi}}]{PhysRevLett.105.032501}%
  \BibitemOpen
  \bibfield  {author} {\bibinfo {author} {\bibfnamefont {T.}~\bibnamefont
  {Otsuka}}, \bibinfo {author} {\bibfnamefont {T.}~\bibnamefont {Suzuki}},
  \bibinfo {author} {\bibfnamefont {J.~D.}\ \bibnamefont {Holt}}, \bibinfo
  {author} {\bibfnamefont {A.}~\bibnamefont {Schwenk}}, \ and\ \bibinfo
  {author} {\bibfnamefont {Y.}~\bibnamefont {Akaishi}},\ }\href {\doibase
  10.1103/PhysRevLett.105.032501} {\bibfield  {journal} {\bibinfo  {journal}
  {Phys. Rev. Lett.}\ }\textbf {\bibinfo {volume} {105}},\ \bibinfo {pages}
  {032501} (\bibinfo {year} {2010})}\BibitemShut {NoStop}%
\bibitem [{\citenamefont {Bogner}\ \emph {et~al.}(2014)\citenamefont {Bogner},
  \citenamefont {Hergert}, \citenamefont {Holt}, \citenamefont {Schwenk},
  \citenamefont {Binder}, \citenamefont {Calci}, \citenamefont {Langhammer},\
  and\ \citenamefont {Roth}}]{PhysRevLett.113.142501}%
  \BibitemOpen
  \bibfield  {author} {\bibinfo {author} {\bibfnamefont {S.~K.}\ \bibnamefont
  {Bogner}}, \bibinfo {author} {\bibfnamefont {H.}~\bibnamefont {Hergert}},
  \bibinfo {author} {\bibfnamefont {J.~D.}\ \bibnamefont {Holt}}, \bibinfo
  {author} {\bibfnamefont {A.}~\bibnamefont {Schwenk}}, \bibinfo {author}
  {\bibfnamefont {S.}~\bibnamefont {Binder}}, \bibinfo {author} {\bibfnamefont
  {A.}~\bibnamefont {Calci}}, \bibinfo {author} {\bibfnamefont
  {J.}~\bibnamefont {Langhammer}}, \ and\ \bibinfo {author} {\bibfnamefont
  {R.}~\bibnamefont {Roth}},\ }\href {\doibase 10.1103/PhysRevLett.113.142501}
  {\bibfield  {journal} {\bibinfo  {journal} {Phys. Rev. Lett.}\ }\textbf
  {\bibinfo {volume} {113}},\ \bibinfo {pages} {142501} (\bibinfo {year}
  {2014})}\BibitemShut {NoStop}%
\bibitem [{\citenamefont {Jansen}\ \emph {et~al.}(2014)\citenamefont {Jansen},
  \citenamefont {Engel}, \citenamefont {Hagen}, \citenamefont {Navratil},\ and\
  \citenamefont {Signoracci}}]{PhysRevLett.113.142502}%
  \BibitemOpen
  \bibfield  {author} {\bibinfo {author} {\bibfnamefont {G.~R.}\ \bibnamefont
  {Jansen}}, \bibinfo {author} {\bibfnamefont {J.}~\bibnamefont {Engel}},
  \bibinfo {author} {\bibfnamefont {G.}~\bibnamefont {Hagen}}, \bibinfo
  {author} {\bibfnamefont {P.}~\bibnamefont {Navratil}}, \ and\ \bibinfo
  {author} {\bibfnamefont {A.}~\bibnamefont {Signoracci}},\ }\href {\doibase
  10.1103/PhysRevLett.113.142502} {\bibfield  {journal} {\bibinfo  {journal}
  {Phys. Rev. Lett.}\ }\textbf {\bibinfo {volume} {113}},\ \bibinfo {pages}
  {142502} (\bibinfo {year} {2014})}\BibitemShut {NoStop}%
\bibitem [{\citenamefont {Holt}\ \emph {et~al.}(2013)\citenamefont {Holt},
  \citenamefont {Men{\'e}ndez},\ and\ \citenamefont {Schwenk}}]{Holt2013}%
  \BibitemOpen
  \bibfield  {author} {\bibinfo {author} {\bibfnamefont {J.~D.}\ \bibnamefont
  {Holt}}, \bibinfo {author} {\bibfnamefont {J.}~\bibnamefont {Men{\'e}ndez}},
  \ and\ \bibinfo {author} {\bibfnamefont {A.}~\bibnamefont {Schwenk}},\ }\href
  {\doibase 10.1140/epja/i2013-13039-2} {\bibfield  {journal} {\bibinfo
  {journal} {Eur. Phys. J. A}\ }\textbf {\bibinfo {volume} {49}},\ \bibinfo
  {pages} {39} (\bibinfo {year} {2013})}\BibitemShut {NoStop}%
\bibitem [{\citenamefont {Stroberg}\ \emph {et~al.}(2017)\citenamefont
  {Stroberg}, \citenamefont {Calci}, \citenamefont {Hergert}, \citenamefont
  {Holt}, \citenamefont {Bogner}, \citenamefont {Roth},\ and\ \citenamefont
  {Schwenk}}]{PhysRevLett.118.032502}%
  \BibitemOpen
  \bibfield  {author} {\bibinfo {author} {\bibfnamefont {S.~R.}\ \bibnamefont
  {Stroberg}}, \bibinfo {author} {\bibfnamefont {A.}~\bibnamefont {Calci}},
  \bibinfo {author} {\bibfnamefont {H.}~\bibnamefont {Hergert}}, \bibinfo
  {author} {\bibfnamefont {J.~D.}\ \bibnamefont {Holt}}, \bibinfo {author}
  {\bibfnamefont {S.~K.}\ \bibnamefont {Bogner}}, \bibinfo {author}
  {\bibfnamefont {R.}~\bibnamefont {Roth}}, \ and\ \bibinfo {author}
  {\bibfnamefont {A.}~\bibnamefont {Schwenk}},\ }\href {\doibase
  10.1103/PhysRevLett.118.032502} {\bibfield  {journal} {\bibinfo  {journal}
  {Phys. Rev. Lett.}\ }\textbf {\bibinfo {volume} {118}},\ \bibinfo {pages}
  {032502} (\bibinfo {year} {2017})}\BibitemShut {NoStop}%
\bibitem [{\citenamefont {Stroberg}\ \emph {et~al.}(2021)\citenamefont
  {Stroberg}, \citenamefont {Holt}, \citenamefont {Schwenk},\ and\
  \citenamefont {Simonis}}]{Holt:2019}%
  \BibitemOpen
  \bibfield  {author} {\bibinfo {author} {\bibfnamefont {S.~R.}\ \bibnamefont
  {Stroberg}}, \bibinfo {author} {\bibfnamefont {J.~D.}\ \bibnamefont {Holt}},
  \bibinfo {author} {\bibfnamefont {A.}~\bibnamefont {Schwenk}}, \ and\
  \bibinfo {author} {\bibfnamefont {J.}~\bibnamefont {Simonis}},\ }\href
  {\doibase 10.1103/PhysRevLett.126.022501} {\bibfield  {journal} {\bibinfo
  {journal} {Phys. Rev. Lett.}\ }\textbf {\bibinfo {volume} {126}},\ \bibinfo
  {pages} {022501} (\bibinfo {year} {2021})}\BibitemShut {NoStop}%
\bibitem [{\citenamefont {Hu}\ \emph {et~al.}(2019)\citenamefont {Hu},
  \citenamefont {Wu}, \citenamefont {Sun},\ and\ \citenamefont
  {Xu}}]{PhysRevC.99.061302}%
  \BibitemOpen
  \bibfield  {author} {\bibinfo {author} {\bibfnamefont {B.~S.}\ \bibnamefont
  {Hu}}, \bibinfo {author} {\bibfnamefont {Q.}~\bibnamefont {Wu}}, \bibinfo
  {author} {\bibfnamefont {Z.~H.}\ \bibnamefont {Sun}}, \ and\ \bibinfo
  {author} {\bibfnamefont {F.~R.}\ \bibnamefont {Xu}},\ }\href {\doibase
  10.1103/PhysRevC.99.061302} {\bibfield  {journal} {\bibinfo  {journal} {Phys.
  Rev. C}\ }\textbf {\bibinfo {volume} {99}},\ \bibinfo {pages} {061302(R)}
  (\bibinfo {year} {2019})}\BibitemShut {NoStop}%
\bibitem [{\citenamefont {Hu}\ \emph {et~al.}(2020)\citenamefont {Hu},
  \citenamefont {Wu}, \citenamefont {Li}, \citenamefont {Ma}, \citenamefont
  {Sun}, \citenamefont {Michel},\ and\ \citenamefont {Xu}}]{HU2020135206}%
  \BibitemOpen
  \bibfield  {author} {\bibinfo {author} {\bibfnamefont {B.~S.}\ \bibnamefont
  {Hu}}, \bibinfo {author} {\bibfnamefont {Q.}~\bibnamefont {Wu}}, \bibinfo
  {author} {\bibfnamefont {J.~G.}\ \bibnamefont {Li}}, \bibinfo {author}
  {\bibfnamefont {Y.~Z.}\ \bibnamefont {Ma}}, \bibinfo {author} {\bibfnamefont
  {Z.~H.}\ \bibnamefont {Sun}}, \bibinfo {author} {\bibfnamefont
  {N.}~\bibnamefont {Michel}}, \ and\ \bibinfo {author} {\bibfnamefont {F.~R.}\
  \bibnamefont {Xu}},\ }\href {\doibase
  https://doi.org/10.1016/j.physletb.2020.135206} {\bibfield  {journal}
  {\bibinfo  {journal} {Phys. Lett. B}\ }\textbf {\bibinfo {volume} {802}},\
  \bibinfo {pages} {135206} (\bibinfo {year} {2020})}\BibitemShut {NoStop}%
\bibitem [{\citenamefont {Ma}\ \emph {et~al.}(2020)\citenamefont {Ma},
  \citenamefont {Xu}, \citenamefont {Coraggio}, \citenamefont {Hu},
  \citenamefont {Li}, \citenamefont {Fukui}, \citenamefont {Angelis},
  \citenamefont {Itaco},\ and\ \citenamefont {Gargano}}]{MA2020135257}%
  \BibitemOpen
  \bibfield  {author} {\bibinfo {author} {\bibfnamefont {Y.~Z.}\ \bibnamefont
  {Ma}}, \bibinfo {author} {\bibfnamefont {F.~R.}\ \bibnamefont {Xu}}, \bibinfo
  {author} {\bibfnamefont {L.}~\bibnamefont {Coraggio}}, \bibinfo {author}
  {\bibfnamefont {B.~S.}\ \bibnamefont {Hu}}, \bibinfo {author} {\bibfnamefont
  {J.~G.}\ \bibnamefont {Li}}, \bibinfo {author} {\bibfnamefont
  {T.}~\bibnamefont {Fukui}}, \bibinfo {author} {\bibfnamefont {L.~D.}\
  \bibnamefont {Angelis}}, \bibinfo {author} {\bibfnamefont {N.}~\bibnamefont
  {Itaco}}, \ and\ \bibinfo {author} {\bibfnamefont {A.}~\bibnamefont
  {Gargano}},\ }\href {\doibase https://doi.org/10.1016/j.physletb.2020.135257}
  {\bibfield  {journal} {\bibinfo  {journal} {Phys. Lett. B}\ }\textbf
  {\bibinfo {volume} {802}},\ \bibinfo {pages} {135257} (\bibinfo {year}
  {2020})}\BibitemShut {NoStop}%
\bibitem [{\citenamefont {Brown}\ and\ \citenamefont
  {Richter}(2006)}]{PhysRevC.74.034315}%
  \BibitemOpen
  \bibfield  {author} {\bibinfo {author} {\bibfnamefont {B.~A.}\ \bibnamefont
  {Brown}}\ and\ \bibinfo {author} {\bibfnamefont {W.~A.}\ \bibnamefont
  {Richter}},\ }\href {\doibase 10.1103/PhysRevC.74.034315} {\bibfield
  {journal} {\bibinfo  {journal} {Phys. Rev. C}\ }\textbf {\bibinfo {volume}
  {74}},\ \bibinfo {pages} {034315} (\bibinfo {year} {2006})}\BibitemShut
  {NoStop}%
\bibitem [{\citenamefont {Yuan}\ \emph {et~al.}(2012)\citenamefont {Yuan},
  \citenamefont {Suzuki}, \citenamefont {Otsuka}, \citenamefont {Xu},\ and\
  \citenamefont {Tsunoda}}]{PhysRevC.85.064324}%
  \BibitemOpen
  \bibfield  {author} {\bibinfo {author} {\bibfnamefont {C.}~\bibnamefont
  {Yuan}}, \bibinfo {author} {\bibfnamefont {T.}~\bibnamefont {Suzuki}},
  \bibinfo {author} {\bibfnamefont {T.}~\bibnamefont {Otsuka}}, \bibinfo
  {author} {\bibfnamefont {F.}~\bibnamefont {Xu}}, \ and\ \bibinfo {author}
  {\bibfnamefont {N.}~\bibnamefont {Tsunoda}},\ }\href {\doibase
  10.1103/PhysRevC.85.064324} {\bibfield  {journal} {\bibinfo  {journal} {Phys.
  Rev. C}\ }\textbf {\bibinfo {volume} {85}},\ \bibinfo {pages} {064324}
  (\bibinfo {year} {2012})}\BibitemShut {NoStop}%
\bibitem [{\citenamefont {Volya}\ and\ \citenamefont
  {Zelevinsky}(2005)}]{PhysRevLett.94.052501}%
  \BibitemOpen
  \bibfield  {author} {\bibinfo {author} {\bibfnamefont {A.}~\bibnamefont
  {Volya}}\ and\ \bibinfo {author} {\bibfnamefont {V.}~\bibnamefont
  {Zelevinsky}},\ }\href {\doibase 10.1103/PhysRevLett.94.052501} {\bibfield
  {journal} {\bibinfo  {journal} {Phys. Rev. Lett.}\ }\textbf {\bibinfo
  {volume} {94}},\ \bibinfo {pages} {052501} (\bibinfo {year}
  {2005})}\BibitemShut {NoStop}%
\bibitem [{\citenamefont {Furutani}\ \emph {et~al.}(1979)\citenamefont
  {Furutani}, \citenamefont {Horiuchi},\ and\ \citenamefont {Tamagaki}}]{FHT1}%
  \BibitemOpen
  \bibfield  {author} {\bibinfo {author} {\bibfnamefont {H.}~\bibnamefont
  {Furutani}}, \bibinfo {author} {\bibfnamefont {H.}~\bibnamefont {Horiuchi}},
  \ and\ \bibinfo {author} {\bibfnamefont {R.}~\bibnamefont {Tamagaki}},\
  }\href {\doibase https://doi.org/10.1143/PTP.62.981} {\bibfield  {journal}
  {\bibinfo  {journal} {Prog. Theor. Phys.}\ }\textbf {\bibinfo {volume}
  {62}},\ \bibinfo {pages} {981 } (\bibinfo {year} {1979})}\BibitemShut
  {NoStop}%
\bibitem [{\citenamefont {Furutani}\ \emph {et~al.}(1980)\citenamefont
  {Furutani}, \citenamefont {Kanada}, \citenamefont {Kaneko}, \citenamefont
  {Nagata}, \citenamefont {Nishioka}, \citenamefont {Okabe}, \citenamefont
  {Saito}, \citenamefont {Sakuda},\ and\ \citenamefont {Seya}}]{FHT2}%
  \BibitemOpen
  \bibfield  {author} {\bibinfo {author} {\bibfnamefont {H.}~\bibnamefont
  {Furutani}}, \bibinfo {author} {\bibfnamefont {H.}~\bibnamefont {Kanada}},
  \bibinfo {author} {\bibfnamefont {T.}~\bibnamefont {Kaneko}}, \bibinfo
  {author} {\bibfnamefont {S.}~\bibnamefont {Nagata}}, \bibinfo {author}
  {\bibfnamefont {H.}~\bibnamefont {Nishioka}}, \bibinfo {author}
  {\bibfnamefont {S.}~\bibnamefont {Okabe}}, \bibinfo {author} {\bibfnamefont
  {S.}~\bibnamefont {Saito}}, \bibinfo {author} {\bibfnamefont
  {T.}~\bibnamefont {Sakuda}}, \ and\ \bibinfo {author} {\bibfnamefont
  {M.}~\bibnamefont {Seya}},\ }\href {\doibase
  https://doi.org/10.1143/PTPS.68.193} {\bibfield  {journal} {\bibinfo
  {journal} {Prog. Theor. Phys. Supp.}\ }\textbf {\bibinfo {volume} {68}},\
  \bibinfo {pages} {193 } (\bibinfo {year} {1980})}\BibitemShut {NoStop}%
\bibitem [{ens()}]{ensdf}%
  \BibitemOpen
  \href@noop {} {}\bibinfo {howpublished}
  {\url{https://www.nndc.bnl.gov/ensdf/}}\BibitemShut {NoStop}%
\bibitem [{\citenamefont {Contessi}\ \emph {et~al.}(2017)\citenamefont
  {Contessi}, \citenamefont {Lovato}, \citenamefont {Pederiva}, \citenamefont
  {Roggero}, \citenamefont {Kirscher},\ and\ \citenamefont {{van
  Kolck}}}]{CONTESSI2017839}%
  \BibitemOpen
  \bibfield  {author} {\bibinfo {author} {\bibfnamefont {L.}~\bibnamefont
  {Contessi}}, \bibinfo {author} {\bibfnamefont {A.}~\bibnamefont {Lovato}},
  \bibinfo {author} {\bibfnamefont {F.}~\bibnamefont {Pederiva}}, \bibinfo
  {author} {\bibfnamefont {A.}~\bibnamefont {Roggero}}, \bibinfo {author}
  {\bibfnamefont {J.}~\bibnamefont {Kirscher}}, \ and\ \bibinfo {author}
  {\bibfnamefont {U.}~\bibnamefont {{van Kolck}}},\ }\href {\doibase
  https://doi.org/10.1016/j.physletb.2017.07.048} {\bibfield  {journal}
  {\bibinfo  {journal} {Phys. Lett. B}\ }\textbf {\bibinfo {volume} {772}},\
  \bibinfo {pages} {839 } (\bibinfo {year} {2017})}\BibitemShut {NoStop}%
\bibitem [{\citenamefont {Hammer}\ \emph {et~al.}(2013)\citenamefont {Hammer},
  \citenamefont {Nogga},\ and\ \citenamefont {Schwenk}}]{RevModPhys.85.197}%
  \BibitemOpen
  \bibfield  {author} {\bibinfo {author} {\bibfnamefont {H.-W.}\ \bibnamefont
  {Hammer}}, \bibinfo {author} {\bibfnamefont {A.}~\bibnamefont {Nogga}}, \
  and\ \bibinfo {author} {\bibfnamefont {A.}~\bibnamefont {Schwenk}},\ }\href
  {\doibase 10.1103/RevModPhys.85.197} {\bibfield  {journal} {\bibinfo
  {journal} {Rev. Mod. Phys.}\ }\textbf {\bibinfo {volume} {85}},\ \bibinfo
  {pages} {197} (\bibinfo {year} {2013})}\BibitemShut {NoStop}%
\bibitem [{\citenamefont {Hammer}\ \emph {et~al.}(2020)\citenamefont {Hammer},
  \citenamefont {K\"onig},\ and\ \citenamefont {van
  Kolck}}]{RevModPhys.92.025004}%
  \BibitemOpen
  \bibfield  {author} {\bibinfo {author} {\bibfnamefont {H.-W.}\ \bibnamefont
  {Hammer}}, \bibinfo {author} {\bibfnamefont {S.}~\bibnamefont {K\"onig}}, \
  and\ \bibinfo {author} {\bibfnamefont {U.}~\bibnamefont {van Kolck}},\ }\href
  {\doibase 10.1103/RevModPhys.92.025004} {\bibfield  {journal} {\bibinfo
  {journal} {Rev. Mod. Phys.}\ }\textbf {\bibinfo {volume} {92}},\ \bibinfo
  {pages} {025004} (\bibinfo {year} {2020})}\BibitemShut {NoStop}%
\bibitem [{\citenamefont {Binder}\ \emph {et~al.}(2016)\citenamefont {Binder},
  \citenamefont {Ekstr\"om}, \citenamefont {Hagen}, \citenamefont
  {Papenbrock},\ and\ \citenamefont {Wendt}}]{PhysRevC.93.044332}%
  \BibitemOpen
  \bibfield  {author} {\bibinfo {author} {\bibfnamefont {S.}~\bibnamefont
  {Binder}}, \bibinfo {author} {\bibfnamefont {A.}~\bibnamefont {Ekstr\"om}},
  \bibinfo {author} {\bibfnamefont {G.}~\bibnamefont {Hagen}}, \bibinfo
  {author} {\bibfnamefont {T.}~\bibnamefont {Papenbrock}}, \ and\ \bibinfo
  {author} {\bibfnamefont {K.~A.}\ \bibnamefont {Wendt}},\ }\href {\doibase
  10.1103/PhysRevC.93.044332} {\bibfield  {journal} {\bibinfo  {journal} {Phys.
  Rev. C}\ }\textbf {\bibinfo {volume} {93}},\ \bibinfo {pages} {044332}
  (\bibinfo {year} {2016})}\BibitemShut {NoStop}%
\bibitem [{\citenamefont {Bansal}\ \emph {et~al.}(2018)\citenamefont {Bansal},
  \citenamefont {Binder}, \citenamefont {Ekstr\"om}, \citenamefont {Hagen},
  \citenamefont {Jansen},\ and\ \citenamefont
  {Papenbrock}}]{PhysRevC.98.054301}%
  \BibitemOpen
  \bibfield  {author} {\bibinfo {author} {\bibfnamefont {A.}~\bibnamefont
  {Bansal}}, \bibinfo {author} {\bibfnamefont {S.}~\bibnamefont {Binder}},
  \bibinfo {author} {\bibfnamefont {A.}~\bibnamefont {Ekstr\"om}}, \bibinfo
  {author} {\bibfnamefont {G.}~\bibnamefont {Hagen}}, \bibinfo {author}
  {\bibfnamefont {G.~R.}\ \bibnamefont {Jansen}}, \ and\ \bibinfo {author}
  {\bibfnamefont {T.}~\bibnamefont {Papenbrock}},\ }\href {\doibase
  10.1103/PhysRevC.98.054301} {\bibfield  {journal} {\bibinfo  {journal} {Phys.
  Rev. C}\ }\textbf {\bibinfo {volume} {98}},\ \bibinfo {pages} {054301}
  (\bibinfo {year} {2018})}\BibitemShut {NoStop}%
\bibitem [{\citenamefont {Furnstahl}\ \emph {et~al.}(2012)\citenamefont
  {Furnstahl}, \citenamefont {Hagen},\ and\ \citenamefont
  {Papenbrock}}]{PhysRevC.86.031301}%
  \BibitemOpen
  \bibfield  {author} {\bibinfo {author} {\bibfnamefont {R.~J.}\ \bibnamefont
  {Furnstahl}}, \bibinfo {author} {\bibfnamefont {G.}~\bibnamefont {Hagen}}, \
  and\ \bibinfo {author} {\bibfnamefont {T.}~\bibnamefont {Papenbrock}},\
  }\href {\doibase 10.1103/PhysRevC.86.031301} {\bibfield  {journal} {\bibinfo
  {journal} {Phys. Rev. C}\ }\textbf {\bibinfo {volume} {86}},\ \bibinfo
  {pages} {031301(R)} (\bibinfo {year} {2012})}\BibitemShut {NoStop}%
\bibitem [{\citenamefont {K\"onig}\ \emph {et~al.}(2014)\citenamefont
  {K\"onig}, \citenamefont {Bogner}, \citenamefont {Furnstahl}, \citenamefont
  {More},\ and\ \citenamefont {Papenbrock}}]{PhysRevC.90.064007}%
  \BibitemOpen
  \bibfield  {author} {\bibinfo {author} {\bibfnamefont {S.}~\bibnamefont
  {K\"onig}}, \bibinfo {author} {\bibfnamefont {S.~K.}\ \bibnamefont {Bogner}},
  \bibinfo {author} {\bibfnamefont {R.~J.}\ \bibnamefont {Furnstahl}}, \bibinfo
  {author} {\bibfnamefont {S.~N.}\ \bibnamefont {More}}, \ and\ \bibinfo
  {author} {\bibfnamefont {T.}~\bibnamefont {Papenbrock}},\ }\href {\doibase
  10.1103/PhysRevC.90.064007} {\bibfield  {journal} {\bibinfo  {journal} {Phys.
  Rev. C}\ }\textbf {\bibinfo {volume} {90}},\ \bibinfo {pages} {064007}
  (\bibinfo {year} {2014})}\BibitemShut {NoStop}%
\bibitem [{\citenamefont {Bohr}\ and\ \citenamefont
  {Mottelson}(1998)}]{Bohr_Mottelson}%
  \BibitemOpen
  \bibfield  {author} {\bibinfo {author} {\bibfnamefont {A.}~\bibnamefont
  {Bohr}}\ and\ \bibinfo {author} {\bibfnamefont {B.}~\bibnamefont
  {Mottelson}},\ }\href@noop {} {\emph {\bibinfo {title} {Nuclear Structure}}}\
  (\bibinfo  {publisher} {World Scientific Publishing Co., Inc, Singapore},\
  \bibinfo {year} {1998})\BibitemShut {NoStop}%
\bibitem [{\citenamefont {Brillouin}(1933)}]{NaturalOrbitals}%
  \BibitemOpen
  \bibfield  {author} {\bibinfo {author} {\bibfnamefont {L.}~\bibnamefont
  {Brillouin}},\ }\href@noop {} {\bibfield  {journal} {\bibinfo  {journal}
  {Act. Sci. Ind.}\ }\textbf {\bibinfo {volume} {71}},\ \bibinfo {pages} {159}
  (\bibinfo {year} {1933})}\BibitemShut {NoStop}%
\bibitem [{\citenamefont {Li}\ \emph {et~al.}(2019)\citenamefont {Li},
  \citenamefont {Michel}, \citenamefont {Hu}, \citenamefont {Zuo},\ and\
  \citenamefont {Xu}}]{PhysRevC.100.054313}%
  \BibitemOpen
  \bibfield  {author} {\bibinfo {author} {\bibfnamefont {J.~G.}\ \bibnamefont
  {Li}}, \bibinfo {author} {\bibfnamefont {N.}~\bibnamefont {Michel}}, \bibinfo
  {author} {\bibfnamefont {B.~S.}\ \bibnamefont {Hu}}, \bibinfo {author}
  {\bibfnamefont {W.}~\bibnamefont {Zuo}}, \ and\ \bibinfo {author}
  {\bibfnamefont {F.~R.}\ \bibnamefont {Xu}},\ }\href {\doibase
  10.1103/PhysRevC.100.054313} {\bibfield  {journal} {\bibinfo  {journal}
  {Phys. Rev. C}\ }\textbf {\bibinfo {volume} {100}},\ \bibinfo {pages}
  {054313} (\bibinfo {year} {2019})}\BibitemShut {NoStop}%
\bibitem [{\citenamefont {Lunderberg}\ \emph {et~al.}(2012)\citenamefont
  {Lunderberg}, \citenamefont {DeYoung}, \citenamefont {Kohley}, \citenamefont
  {Attanayake}, \citenamefont {Baumann}, \citenamefont {Bazin}, \citenamefont
  {Christian}, \citenamefont {Divaratne}, \citenamefont {Grimes}, \citenamefont
  {Haagsma}, \citenamefont {Finck}, \citenamefont {Frank}, \citenamefont
  {Luther}, \citenamefont {Mosby}, \citenamefont {Nagi}, \citenamefont
  {Peaslee}, \citenamefont {Schiller}, \citenamefont {Snyder}, \citenamefont
  {Spyrou}, \citenamefont {Strongman},\ and\ \citenamefont
  {Thoennessen}}]{PhysRevLett.108.142503}%
  \BibitemOpen
  \bibfield  {author} {\bibinfo {author} {\bibfnamefont {E.}~\bibnamefont
  {Lunderberg}}, \bibinfo {author} {\bibfnamefont {P.~A.}\ \bibnamefont
  {DeYoung}}, \bibinfo {author} {\bibfnamefont {Z.}~\bibnamefont {Kohley}},
  \bibinfo {author} {\bibfnamefont {H.}~\bibnamefont {Attanayake}}, \bibinfo
  {author} {\bibfnamefont {T.}~\bibnamefont {Baumann}}, \bibinfo {author}
  {\bibfnamefont {D.}~\bibnamefont {Bazin}}, \bibinfo {author} {\bibfnamefont
  {G.}~\bibnamefont {Christian}}, \bibinfo {author} {\bibfnamefont
  {D.}~\bibnamefont {Divaratne}}, \bibinfo {author} {\bibfnamefont {S.~M.}\
  \bibnamefont {Grimes}}, \bibinfo {author} {\bibfnamefont {A.}~\bibnamefont
  {Haagsma}}, \bibinfo {author} {\bibfnamefont {J.~E.}\ \bibnamefont {Finck}},
  \bibinfo {author} {\bibfnamefont {N.}~\bibnamefont {Frank}}, \bibinfo
  {author} {\bibfnamefont {B.}~\bibnamefont {Luther}}, \bibinfo {author}
  {\bibfnamefont {S.}~\bibnamefont {Mosby}}, \bibinfo {author} {\bibfnamefont
  {T.}~\bibnamefont {Nagi}}, \bibinfo {author} {\bibfnamefont {G.~F.}\
  \bibnamefont {Peaslee}}, \bibinfo {author} {\bibfnamefont {A.}~\bibnamefont
  {Schiller}}, \bibinfo {author} {\bibfnamefont {J.}~\bibnamefont {Snyder}},
  \bibinfo {author} {\bibfnamefont {A.}~\bibnamefont {Spyrou}}, \bibinfo
  {author} {\bibfnamefont {M.~J.}\ \bibnamefont {Strongman}}, \ and\ \bibinfo
  {author} {\bibfnamefont {M.}~\bibnamefont {Thoennessen}},\ }\href {\doibase
  10.1103/PhysRevLett.108.142503} {\bibfield  {journal} {\bibinfo  {journal}
  {Phys. Rev. Lett.}\ }\textbf {\bibinfo {volume} {108}},\ \bibinfo {pages}
  {142503} (\bibinfo {year} {2012})}\BibitemShut {NoStop}%
\bibitem [{\citenamefont {Wang}\ \emph {et~al.}(2017)\citenamefont {Wang},
  \citenamefont {Audi}, \citenamefont {Kondev}, \citenamefont {Huang},
  \citenamefont {Naimi},\ and\ \citenamefont {Xu}}]{wang2017ame2016}%
  \BibitemOpen
  \bibfield  {author} {\bibinfo {author} {\bibfnamefont {M.}~\bibnamefont
  {Wang}}, \bibinfo {author} {\bibfnamefont {G.}~\bibnamefont {Audi}}, \bibinfo
  {author} {\bibfnamefont {F.}~\bibnamefont {Kondev}}, \bibinfo {author}
  {\bibfnamefont {W.}~\bibnamefont {Huang}}, \bibinfo {author} {\bibfnamefont
  {S.}~\bibnamefont {Naimi}}, \ and\ \bibinfo {author} {\bibfnamefont
  {X.}~\bibnamefont {Xu}},\ }\href@noop {} {\bibfield  {journal} {\bibinfo
  {journal} {Chin. Phys. C}\ }\textbf {\bibinfo {volume} {41}},\ \bibinfo
  {pages} {030003} (\bibinfo {year} {2017})}\BibitemShut {NoStop}%
\bibitem [{\citenamefont {Grigorenko}\ \emph {et~al.}(2011)\citenamefont
  {Grigorenko}, \citenamefont {Mukha}, \citenamefont {Scheidenberger},\ and\
  \citenamefont {Zhukov}}]{PhysRevC.84.021303}%
  \BibitemOpen
  \bibfield  {author} {\bibinfo {author} {\bibfnamefont {L.~V.}\ \bibnamefont
  {Grigorenko}}, \bibinfo {author} {\bibfnamefont {I.~G.}\ \bibnamefont
  {Mukha}}, \bibinfo {author} {\bibfnamefont {C.}~\bibnamefont
  {Scheidenberger}}, \ and\ \bibinfo {author} {\bibfnamefont {M.~V.}\
  \bibnamefont {Zhukov}},\ }\href {\doibase 10.1103/PhysRevC.84.021303}
  {\bibfield  {journal} {\bibinfo  {journal} {Phys. Rev. C}\ }\textbf {\bibinfo
  {volume} {84}},\ \bibinfo {pages} {021303(R)} (\bibinfo {year}
  {2011})}\BibitemShut {NoStop}%
\end{thebibliography}%

\end{document}